\documentclass[twocolumn]{aastex63}

\usepackage{threeparttable,hyperref}
\usepackage{amsmath,physics}

\newcommand{\nsplash}{6405}

\newcommand{\fkickedupdiskdorman}{5.2\% $\pm$ 1.2\%}
\newcommand{\fmrtot}{4.6\%}
\newcommand{\fkickedupdiskexcess}{$5.7\% \pm 0.2\%$} 
\newcommand{\fdisktotstat}{$74.9\% \pm 4.5$\%} 

\newcommand{\fehphotshiftfourtotwelve}{$-0.26$}
\newcommand{\medfehphotneshelf}{$-0.16$}
\newcommand{\medfehphotminhalo}{$-0.15$}
\newcommand{\fehphotshiftnoinnerfield}{$-0.29$}
\newcommand{\frachostnbody}{98.9\%}

\newcommand{\ndiskrgb}{2500} 
\newcommand{\nhalorgb}{554} 
\newcommand{\nmixedrgb}{458} 

\newcommand{\fehphotdiskrgbmr}{$-0.10$}
\newcommand{\fehphotdiskrgbmp}{$-0.90$} 
\newcommand{\fehphotdiskrgbmrsig}{$0.23$}
\newcommand{\fehphotdiskrgbmpsig}{$0.54$} 
\newcommand{\fehphotdiskrgbmrfrac}{71\%}
\newcommand{\fehphotdiskrgbmpfrac}{29\%}

\newcommand{\fehphothalorgbmr}{$-0.18$} 
\newcommand{\fehphothalorgbmp}{$-0.89$} 
\newcommand{\fehphothalorgbmrsig}{$0.22$}
\newcommand{\fehphothalorgbmpsig}{$0.44$} 
\newcommand{\fehphothalorgbmrfrac}{48\%} 
\newcommand{\fehphothalorgbmpfrac}{52\%} 

\newcommand{\fcontamhalo}{7.4\%}
\newcommand{\fcontamdisk}{7.8\%}

\newcommand{\fehphotmedhalorgbcontamcorr}{$-0.50^{+0.003}_{-0.002}$}
\newcommand{\fehphotmeddiskrgbcontamcorr}{$-0.17^{+0.001}_{-0.003}$}
\newcommand{\fehphotavghalorgbcontamcorr}{$-0.58 \pm 0.002$}
\newcommand{\fehphotavgdiskrgbcontamcorr}{$-0.31 \pm 0.002$}

\newcommand{\gradientdisk}{$-0.0176$}
\newcommand{\gradientdiskerr}{0.0002}
\newcommand{\gradienthalo}{$-0.0029$}
\newcommand{\gradienthaloerr}{0.0004}
\newcommand{\interceptdisk}{+0.037}
\newcommand{\interceptdiskerr}{0.004}
\newcommand{\intercepthalo}{$-0.335$}
\newcommand{\intercepthaloerr}{0.005}

\newcommand{\gradientdisklowext}{$-0.0200$}
\newcommand{\gradientdisklowexterr}{0.0003}
\newcommand{\gradienthalolowext}{$-0.0031$}
\newcommand{\gradienthalolowexterr}{0.0005}
\newcommand{\interceptdisklowext}{+0.092}
\newcommand{\interceptdisklowexterr}{0.005}
\newcommand{\intercepthalolowext}{$-0.318$}
\newcommand{\intercepthalolowexterr}{$0.007$}

\newcommand{\gradientdiskdust}{$-0.0200$}
\newcommand{\gradientdiskdusterr}{0.0002}
\newcommand{\gradienthalodust}{$-0.0055$}
\newcommand{\gradienthalodusterr}{0.0004}
\newcommand{\interceptdiskdust}{+0.092}
\newcommand{\interceptdiskdusterr}{0.004}
\newcommand{\intercepthalodust}{$-0.269$}
\newcommand{\intercepthalodusterr}{$0.005$}

\newcommand{\gradientdiskfid}{$-0.018^{+0.003}_{-0.001}$}
\newcommand{\gradienthalofid}{$-0.003^{+0.003}_{-0.001}$}

\newcommand{\fracmrhalofid}{52.9\%} 
\newcommand{\fracmrhalocontamcorr}{49.3\%} 
\newcommand{\fracmrhalocontambydisk}{3.6\%} 
\newcommand{\fracpkickfid}{28.9\%} 
\newcommand{\fracpkickcontamcorr}{26.7\%} 
\newcommand{\fracpkickcontam}{2.2\%} 

\newcommand{\muvhalomr}{$-280.9$}
\newcommand{\muverrhalomr}{6.1}
\newcommand{\muxhalomr}{$-0.15$}
\newcommand{\muxerrhalomr}{0.03}
\newcommand{\sigvhalomr}{42.9}
\newcommand{\sigverrhalomr}{4.9}
\newcommand{\sigxhalomr}{0.20}
\newcommand{\sigxerrhalomr}{0.02}
\newcommand{\rhalomr}{0.25}
\newcommand{\rerrhalomr}{0.12}
\newcommand{\fmrhalo}{0.26}
\newcommand{\fmrerrhalo}{0.03}

\newcommand{\muvhalomp}{$-347.1$}
\newcommand{\muverrhalomp}{7.4}
\newcommand{\sigvhalomp}{141.0}
\newcommand{\sigverrhalomp}{4.9}
\newcommand{\muxhalomp}{$-0.69$}
\newcommand{\muxerrhalomp}{0.03}
\newcommand{\sigxhalomp}{0.50}
\newcommand{\sigxerrhalomp}{0.02}
\newcommand{\rhalomp}{0.04}
\newcommand{\rerrhalomp}{0.05}

\newcommand{\admeddiskall}{$61.0^{+0.9}_{-0.8}$}
\newcommand{\admedmixedall}{$57.6^{+3.5}_{-2.6}$}
\newcommand{\admedhaloall}{$66.4^{+1.5}_{-2.2}$}
\newcommand{\admedhalomrall}{$63.8^{+2.9}_{-1.2}$}
\newcommand{\admedhalompall}{$67.9^{+1.0}_{-2.0}$}

\newcommand{\percentmrhalo}{71.8\%}
\newcommand{\percentmrdisk}{79.5\%}
\newcommand{\percentmrmixed}{84.7\%}

\newcommand{\admeddisk}{$68.7^{+0.9}_{-0.8}$}
\newcommand{\admedmixed}{$75.1^{+3.2}_{-1.4}$}
\newcommand{\admedhalo}{$74.6^{+1.4}_{-1.1}$}
\newcommand{\admedhalomr}{$77.4^{+0.9}_{-3.6}$}
\newcommand{\admedhalomp}{$73.4^{+1.6}_{-4.6}$}

\newcommand{\addiskhalodiffsig}{2.9}
\newcommand{\addiskmixeddiffsig}{2.7}
\newcommand{\adhalomrdiffsigdiskall}{1.3}
\newcommand{\adhalomrdiffsigdiskonaxis}{1.9}
\newcommand{\adhalomrdiffsigrest}{1}
\newcommand{\addiskhalodiffsigall}{1.8}

\newcommand{\fehphotmeddiskrgb}{$-0.19$}
\newcommand{\fehphotmedmixedrgb}{$-0.16$}
\newcommand{\fehphotmedhalorgb}{$-0.46$} 

\newcommand{\meddeltaewna}{2.5}
\newcommand{\nmembersplash}{4844}
\newcommand{\meanewand}{0.54}
\newcommand{\stdewand}{1.00}
\newcommand{\meanewmw}{2.27}
\newcommand{\stdewmw}{1.96}
\newcommand{\percentnonasplash}{51.6\%}
\newcommand{\nremovecmd}{1443}
\newcommand{\nremovena}{118}
\newcommand{\fracgiantnoew}{43.6\%}
\newcommand{\fracrgbnowew}{34.9\%}

\newcommand{\fmwcontam}{0.2\%}

\newcommand{\muhaloone}{$-$258.2}
\newcommand{\muhalotwo}{$-$268.7}
\newcommand{\muhalothree}{$-$238.8}
\newcommand{\sighaloone}{134.4}
\newcommand{\sighalotwo}{135.3}
\newcommand{\sighalothree}{117.5}
\newcommand{\percentdisksplash}{72.6}
\newcommand{\percenthalosplash}{14.4}
\newcommand{\percentmixedsplash}{13.0}

\newcommand{\medradneone}{6.3}
\newcommand{\medradnetwo}{10.3}
\newcommand{\medradnethree}{14.4}

\newcommand{\fehphotmedallrgb}{$-0.18$}
\newcommand{\fehphotmedlowextrgb}{$-0.21$}
\newcommand{\fehphotdiffrgbcorr}{$-0.04$}
\newcommand{\sigmaavmed}{0.30}

\newcommand{\diskscaleheight}{0.77}

\newcommand{\nrgblowext}{2054} 
\newcommand{\nrgbdustcorr}{3512} 

\newcommand{\fehphoterrmed}{0.03}

\newcommand{\medrederr}{0.004}
\newcommand{\medblueerr}{0.015}

\newcommand{\difffehphotrgb}{$-0.22$}

\newcommand{\nrgbsplashmwcorr}{3874} 

\newcommand{\ewna}{EW$_{\rm Na}$}
\newcommand{\fehphot}{[Fe/H]$_{\rm phot}$}

\newcommand{\feh}{[Fe/H]}

\newcommand{\vhelio}{$v_{\rm helio}$}
\newcommand{\rproj}{R$_{\rm proj}$}
\newcommand{\fehphotinit}{[Fe/H]$_{\rm phot, init}$}
\newcommand{\fehphoterr}{$\delta$[Fe/H]$_{\rm phot}$}

\newcommand{\avdal}{$A_{V, {\rm Dal}}$}
\newcommand{\sigav}{$\sigma_V$}
\newcommand{\fred}{$f_{\rm red}$}
\newcommand{\pdisk}{$p_{\rm disk}$}

\newcommand{\kms}{km s$^{-1}$}

\received{Sep 16 2022}
\revised{Nov 21 2022}
\accepted{Dec 6 2022}
\submitjournal{AJ}

\shorttitle{Chemodynamics of M31's Major-Axis Halo}
\shortauthors{Escala et al.}

\graphicspath{{./}{figures/}}
\turnoffedit

\begin{document}


\title{Resolved SPLASH Chemodynamics in Andromeda's PHAT Stellar Halo and Disk: On the Nature of the Inner Halo Along the Major Axis}


\correspondingauthor{I. Escala}
\email{iescala@carnegiescience.edu; iescala@princeton.edu}

\author[0000-0002-9933-9551]{Ivanna Escala}
\altaffiliation{Carnegie-Princeton Fellow}
\affiliation{Department of Astrophysical Sciences, Princeton University, 4 Ivy Lane, Princeton, NJ 08544, USA}
\affiliation{The Observatories of the Carnegie Institution for Science, 813 Santa Barbara St, Pasadena, CA 91101, USA}

\author[0000-0001-8481-2660]{Amanda C.~N. Quirk}
\affiliation{Department of Astronomy, Columbia University, 538 West 120th Street, New York, NY 10027, USA}
\affiliation{UCO/Lick Observatory, Department of Astronomy \& Astrophysics, University of California Santa Cruz, 1156 High Street, Santa Cruz, CA 95064, USA}

\author[0000-0001-8867-4234]{Puragra Guhathakurta}
\affiliation{UCO/Lick Observatory, Department of Astronomy \& Astrophysics, University of California Santa Cruz, 1156 High Street, Santa Cruz, CA 95064, USA}

\author[0000-0003-0394-8377]{Karoline M. Gilbert}
\affiliation{Space Telescope Science Institute, 3700 San Martin Drive, Baltimore, MD 21218, USA}
\affiliation{The William H. Miller III Department of Physics \& Astronomy, Bloomberg Center for Physics and Astronomy, Johns Hopkins University, 3400 N. Charles Street, Baltimore, MD 21218}


\author[0000-0002-3233-3032]{J. Leigh Wojno}
\affiliation{Max Planck Institute for Astronomy, Königstuhl 17, 69117 Heidelberg, Germany}
\affiliation{The William H. Miller III Department of Physics \& Astronomy, Bloomberg Center for Physics and Astronomy, Johns Hopkins University, 3400 N. Charles Street, Baltimore, MD 21218}

\author[0000-0001-8536-0547]{Lara Cullinane}
\affiliation{The William H. Miller III Department of Physics \& Astronomy, Bloomberg Center for Physics and Astronomy, Johns Hopkins University, 3400 N. Charles Street, Baltimore, MD 21218}

\author[0000-0002-7502-0597]{Benjamin F. Williams}
\affiliation{Department of Astronomy, University of Washington, Box 351580, Seattle, WA 98195, USA}

\author[0000-0002-1264-2006]{Julianne Dalcanton}
\affiliation{Center for Computational Astrophysics, Flatiron Institute, 162 5th Avenue, New York, NY 10010, USA}
\affiliation{Department of Astronomy, University of Washington, Box 351580, Seattle, WA 98195, USA}

\begin{abstract}
Stellar kinematics and metallicity are key to exploring formation scenarios for galactic disks and halos. In this work, we characterized the relationship between kinematics and photometric metallicity 
along the line-of-sight to M31's disk.
We combined optical HST/ACS photometry from the Panchromatic Hubble Andromeda Treasury (PHAT) survey with Keck/DEIMOS 
spectra
from the Spectroscopic and Photometric Landscape of Andromeda's Stellar Halo (SPLASH) survey. The resulting sample of \nrgbdustcorr\ individual red giant branch stars spans 4--19 projected kpc, making it a useful probe of both the disk and inner halo.
We separated these stars into disk and halo populations by modeling the line-of-sight velocity distributions as a function of position across the disk region, where $\sim$73\% stars have a high likelihood of belonging to the disk and $\sim$14\% to the halo. Although stellar halos are typically thought to be metal-poor, the kinematically identified halo contains a significant population of stars ($\sim$29\%)
with disk-like metallicity (\fehphot\ $\sim$ \fehphotdiskrgbmr). 
This metal-rich halo population lags the gaseous disk to a similar extent as the rest of the halo, indicating that it does not correspond to a canonical thick disk. Its 
properties are inconsistent with those of tidal debris originating from the Giant Stellar Stream merger event. 
Moreover, the halo is chemically distinct from the phase-mixed component previously identified along the minor axis (i.e., away from the disk), implying contributions from different formation channels. These metal-rich halo stars provide direct chemodynamical 
evidence in favor of the previously suggested ``kicked-up'' disk population in M31's inner stellar halo.
\end{abstract}

\keywords{Galaxy stellar disks (1594), Galaxy stellar halos (598), Andromeda Galaxy (39), Galaxy formation (595), Galaxy stellar content (621), Stellar kinematics (1608), Stellar abundances (1577)}

\section{Introduction} \label{sec:intro}

The formation and evolution of stellar disks is a significant part of the mass assembly history of galaxies. In the local universe, disk galaxies have been observed to possess dynamically hot ``thick'' disk components in their vertical structure (e.g., \citealt{ChibaBeers2000,YoachimDalcanton2006}; but see also \citealt{Bovy2012}). Multiple physical mechanisms have been proposed to explain the formation of thick disks, including early in-situ star formation within a turbulent gaseous disk \citep{Bournard2009,Forbes2012} and heating of an initially thin stellar disk by internal perturbations (e.g., \citealt{SellwoodCarlberg1984,JenkinsBinney1990,Lacey1984,SB09a,SB09b,Loebman2011}). However, the hierarchical assembly process may also provide channels for thick disk formation via the accretion of gas (e.g., \citealt{Brook2004}) and ex-situ stellar material \citep{Abadi2003} deposited at large scale heights or heating driven by satellite impacts \citep{Quinn1993,VelazquezWhite1999,Hopkins2008,Kazantzidis2008,VillalobosHelmi2008,Purcell2009}. The formation of the dynamically hot thick disk may therefore be inextricably connected to the inner stellar halo by merger history.

M31's proximity (785 kpc) and inclination (77$^\circ$) result in an unrivaled opportunity to study galaxy disks beyond the Milky Way (MW).
Resolved stellar spectroscopy along the line-of-sight to M31's disk has transformed our understanding of the galaxy's inner structure. Based on spectra of over 5000 red giant branch (RGB) stars, \citet{Dorman2012} discovered a rotating spheroid ($\sigma_v$ $\sim$ 140 \kms) that exceeded $\sim$10\% of the total stellar population in a region dominated by the disk in UV/optical images ($R_{\rm proj}$ $\sim$ 4--20 kpc).
However, surface-brightness decompositions indicated that the bulge should not contribute to stellar populations outside of M31's inner few kiloparsecs \citep{Courteau2011,Gilbert2012,Williams2012}. The structural decomposition of \citet{Dorman2013}, which simultaneously accounted for 
surface brightness profiles, the luminosity function, and kinematics, demonstrated that the spheroid corresponded to a stellar halo. 
\citet{Dorman2013} also found evidence for an excess of stars with a disk-like luminosity function compared to dynamically-based expectations for the disk contribution. These stars could have been born in the disk, but kinematically heated into the stellar halo.

Further studies imply a possible merger origin for the connection between the formation of the halo and disk. \citet{Dorman2015} revealed a steep relationship between age and velocity dispersion in M31's disk region ($\sigma_v$ $\sim$ 90 \kms\ for 4 Gyr ages), concluding that this could be explained by a combination of continuous but non-uniform heating of the disk by mergers. \citet{Quirk2019} identified a trend of monotonically increasing asymmetric drift with stellar age that was consistent with a 4:1 merger event occurring within the last 4 Gyr \citep{QuirkPatel2020}. Moreover, a recent major merger (as explored by \citealt{Hammer2018,DSouzaBell2018}) could simultaneously explain M31's 2--4 Gyr old global burst of star formation \citep{Bernard2015a,Williams2015}, its chemically homogeneous extended disk (15--40 kpc; \citealt{Ibata2005}), and thickened disk structure and kinematics (\citealt{Dorman2012,Dorman2015,Dalcanton2015}; but see also \citealt{Collins2011}). Studies of planetary nebulae in M31's disk have similarly concluded that its age-dispersion relation could result from a major merger \citep{Bhattacharya2019}.

Despite the wealth of kinematical information, comprehensive {\it chemodynamical} investigations are lacking across M31's disk region. Prior studies of resolved stellar populations have been largely restricted to either chemical analyses informed by photometry (e.g., \citealt{Gregersen2015,Telford2019}) or focused on dynamics (e.g., \citealt{Dorman2012,Dorman2013,Dorman2015,Quirk2019}). In order to circumvent crowding, previous chemodynamical efforts have been limited to the outer disk ($\gtrsim$ 15 kpc) in the northeast \citep{Ibata2005} and the southwest \citep{Collins2011}. To date, chemical abundances ([Fe/H] and [$\alpha$/Fe]) for individual stars have been measured for only a small sample of stars in M31's outer disk (at 26 projected kpc; \citealt{Escala2020a}). Recently, oxygen and argon abundances from emission lines in planetary nebulae have provided an additional method to probe the chemical evolutionary history of M31's inner disk \citep{Bhattacharya2022,Arnaboldi2022}.

In this work, we combine resolved spectroscopy from the Spectroscopic and Photometric Landscape of Andromeda's Stellar Halo (SPLASH; \citealt{Guhathakurta2005,Gilbert2006}) survey and photometry from the Panchromatic Hubble Andromeda Treasury (PHAT; \citealt{Dalcanton2012,Williams2014}) to perform the first large-scale stellar chemodynamical analysis of M31's inner disk region ($\lesssim$ 15 kpc) in an effort to explore disk-halo formation scenarios. This approach enables us to disentangle the inner stellar halo from the disk as well as identify chemically distinct stellar populations. In Section~\ref{sec:data}, we introduce the photometric and spectroscopic data sets used in this work. We evaluate M31 membership and perform a kinematical decomposition of the disk region in Section~\ref{sec:sec3}. We correct for dust effects and describe photometric metallicity measurements for RGB stars in Section~\ref{sec:phot_params}. Section~\ref{sec:results} investigates the chemical and dynamical properties of the disk and halo, whereas Section~\ref{sec:discuss} places these results in the context of the literature on M31, the MW, and disk galaxy formation in general. We summarize our main findings in Section~\ref{sec:summary}.

\section{Data}
\label{sec:data}

\subsection{PHAT Photometry} \label{sec:phot}

We used stellar catalogs based on Hubble Space Telescope (HST) Wide Field Camera 3 (WFC3) and Advanced Camera for Surveys (ACS) images from the PHAT survey \citep{Dalcanton2012,Williams2014}). PHAT produced six-filter UV (F275W, F336W), optical (F475W, F814W), and IR (F110W, F160W) photometry across the northeastern disk of M31 out to 20 projected kpc from M31's center for 117 million stars. In particular, we used second generation PHAT photometry, in contrast to \citet{Dorman2012,Dorman2013,Dorman2015} (Section~\ref{sec:splash}). The primary difference between the second \citep{Williams2014} and first \citep{Dalcanton2012} generation photometry is the simultaneous use of all six HST filters for source identification and point-spread-function fitting, which enables significant increases in the completeness-limited photometric depth (F475W $\sim$ 28 in the outer disk) and photometric and astrometric accuracy ($<$5--10 mas). We matched the right ascension and declination (based on v1 PHAT photometry) of SPLASH stars to the updated positions in the PHAT v2 catalog. We searched for matches within a 2 arcsec on-a-side box centered on the v1 astrometry, without applying shifts or offsets between the astrometric versions. If there were multiple matches based on this criterion, we additionally matched by optical photometry within the 2$\sigma$ v2 uncertainties in the F475W and F814W filters (median \medblueerr\ and \medrederr\ mag respectively).

\subsection{SPLASH Spectroscopy} \label{sec:splash}

\begin{figure}
    \centering
    \includegraphics[width=\columnwidth]{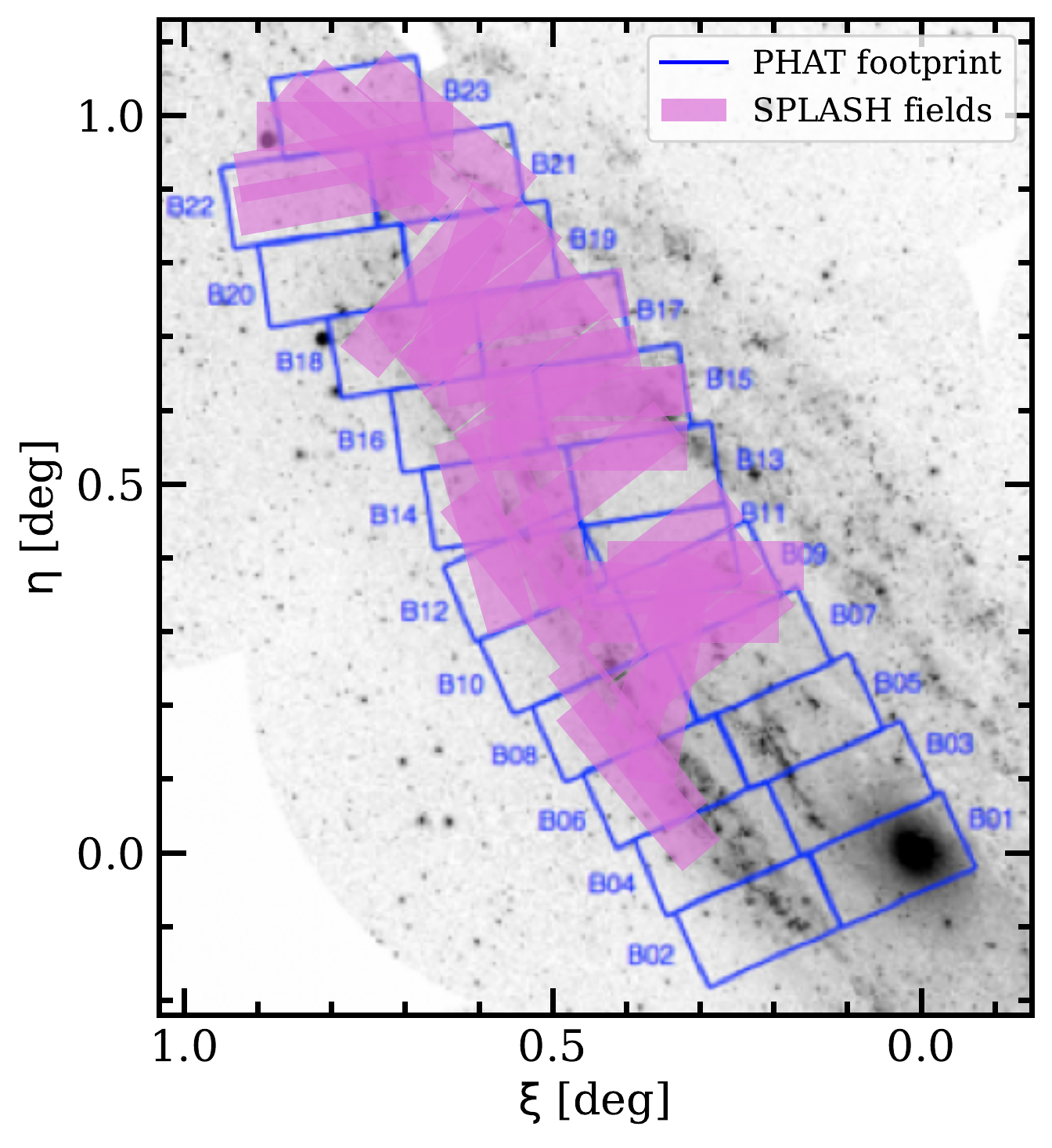}
    \caption{Location of DEIMOS fields from SPLASH (magenta patches; Section~\ref{sec:splash}; \citealt{Dorman2012,Dorman2013,Dorman2015}) in M31-centric coordinates. The approximate size and orientation of each field is represented by a 16' $\times$ 4' rectangle. The outline of the PHAT footprint (Section~\ref{sec:phot}; \citealt{Dalcanton2012}) is shown as blue lines. 
    The NUV image of M31 is from the PHAT archive and was taken by GALEX.}
    \label{fig:fields}
\end{figure}

The SPLASH survey collected $\sim$10,000 Keck/DEIMOS \citep{Faber2003} spectra across M31's northeastern disk to investigate its line-of-sight velocity distribution and stellar properties \citep{Dorman2012,Dorman2013,Dorman2015}. SPLASH stars were targeted based on a mixture of Canada-France-Hawaii Telescope (CFHT) MegaCam photometry and PHAT v1 photometry (Section~\ref{sec:phot}). Each slitmask was observed for $\sim$1 hr using the 1200 $\ell$/mm (pre-2012 observations; \citealt{Dorman2012,Dorman2013}) or 600 $\ell$/mm (post-2012 observations; \citealt{Dorman2015}) grating on DEIMOS. \edit1{The 1200 (600) $\ell$/mm grating configuration results in a spectral resolution of $R \sim 6000$ ($R \sim 3000)$ at the Ca II triplet ($\lambda\lambda \sim 8500$) and a wavelength coverage of $\sim$6300--9100 (4500--9100) \AA\ when combined with the OG550 (GG455) order blocking filter.}

\edit1{The two-dimensional and one-dimensional spectra were reduced and extracted using the {\tt spec2d} and {\tt spec1d} pipelines \citep{Cooper2012,Newman2013}.}
\citet{Dorman2012,Dorman2013,Dorman2015} measured radial velocities from SPLASH spectra using the cross-correlation technique of \citet{SimonGeha2007} including A-band corrections for slit miscentering \citep{Sohn2007} and heliocentric corrections. 
\edit1{To assess the reliability of the velocity measurements, the raw two-dimensional spectra, extracted one-dimensional spectra, and best-fit empirical templates were visually inspected in the {\it zspec} software (D.~Madgwick, DEEP2 survey) and assigned a quality code ($Q$; e.g., \citealt{Guhathakurta2006}). Velocity measurements deemed successful ($Q = 3$ or $Q = 4$) are those that were based on at least one strong spectral feature.}
The typical statistical velocity uncertainty derived from the cross-correlation is roughly a few \kms\ ($\sim$ 10 \kms) for spectra obtained with the 1200 (600) $\ell$/mm grating. 


\begin{figure*}
    \centering
    \includegraphics[width=0.8\textwidth]{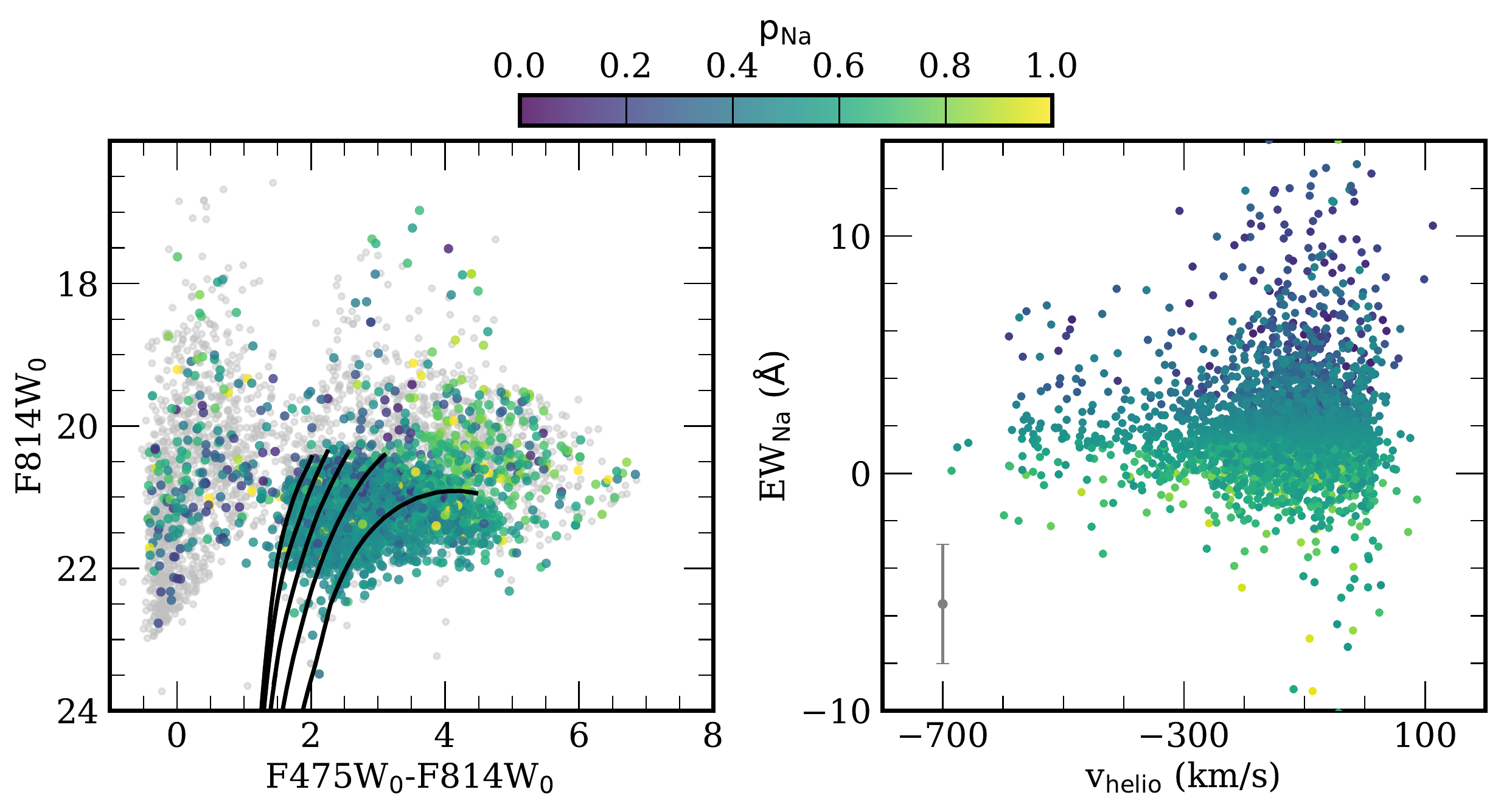}
    \caption{
    Membership determination for SPLASH targets with successful radial velocity measurements \citep{Dorman2012,Dorman2013,Dorman2015} along the line-of-sight to M31's disk (Section~\ref{sec:members}). (Left) Foreground-extinction-corrected (F475W, F814W) PHAT v2 CMD (\citealt{Dalcanton2012,Williams2014}; Section~\ref{sec:phot},~\ref{sec:zinit}) for stars with (colored points) and without (grey points) Na I $\lambda$8190 doublet equivalent width (\ewna) measurements. We show \edit1{4} Gyr PARSEC RGB isochrones \citep{Marigo2017} with \fehphot\ = $-2.0, -1.5, -1.0, -0.5$, 0 for reference, assuming $m-M = 24.45 \pm 0.05$ \citep{Dalcanton2012}. 
    Stars with $p_{\rm Na} \leq 0.25$, which are $\geq$3 times more likely to belong to the MW foreground than M31, are classified as non-members. We also exclude stars with colors bluer than the most metal-poor RGB isochrone. (Right) \ewna\ versus heliocentric velocity (\vhelio). 
    We show the median uncertainty in \ewna\ ($\delta$\ewna\ = \meddeltaewna\ \AA) as a gray errorbar. 
    Stars with $p_{\rm Na} \leq 0.25$ preferentially have blue colors or lie above the TRGB and have more positive \vhelio, all of which are properties characteristic of MW foreground dwarfs.
    }
    \label{fig:members}
\end{figure*}

 Figure~\ref{fig:fields} shows the location of SPLASH fields within the PHAT footprint. The left panel of Figure~\ref{fig:members} shows the (F475W$_0$, F814W$_0$) color-magnitude diagram (CMD; Section~\ref{sec:phot_params}) of \nsplash\ stars in the SPLASH survey of M31's disk with PHAT v2 photometry (Section~\ref{sec:phot}) and successful radial velocity measurements. We also show \edit1{4} Gyr PARSEC RGB isochrones \citep{Marigo2017} spanning $-2 <$ \fehphot\ $< 0$, assuming $m - M$ = 24.45 $\pm$ 0.05 as in \citet{Dalcanton2012}. SPLASH stars span various evolutionary stages, including the upper main sequence of M31's disk, the horizontal branch, the intermediate-age 
 asymptotic giant branch (AGB), and the old RGB, where contamination from the MW foreground is concentrated at observed colors of F475W--F814W $\lesssim$ 2 \citep{Dorman2015}.

\section{Membership and Kinematics}
\label{sec:sec3}

\begin{figure}
    \centering
    \includegraphics[width=\columnwidth]{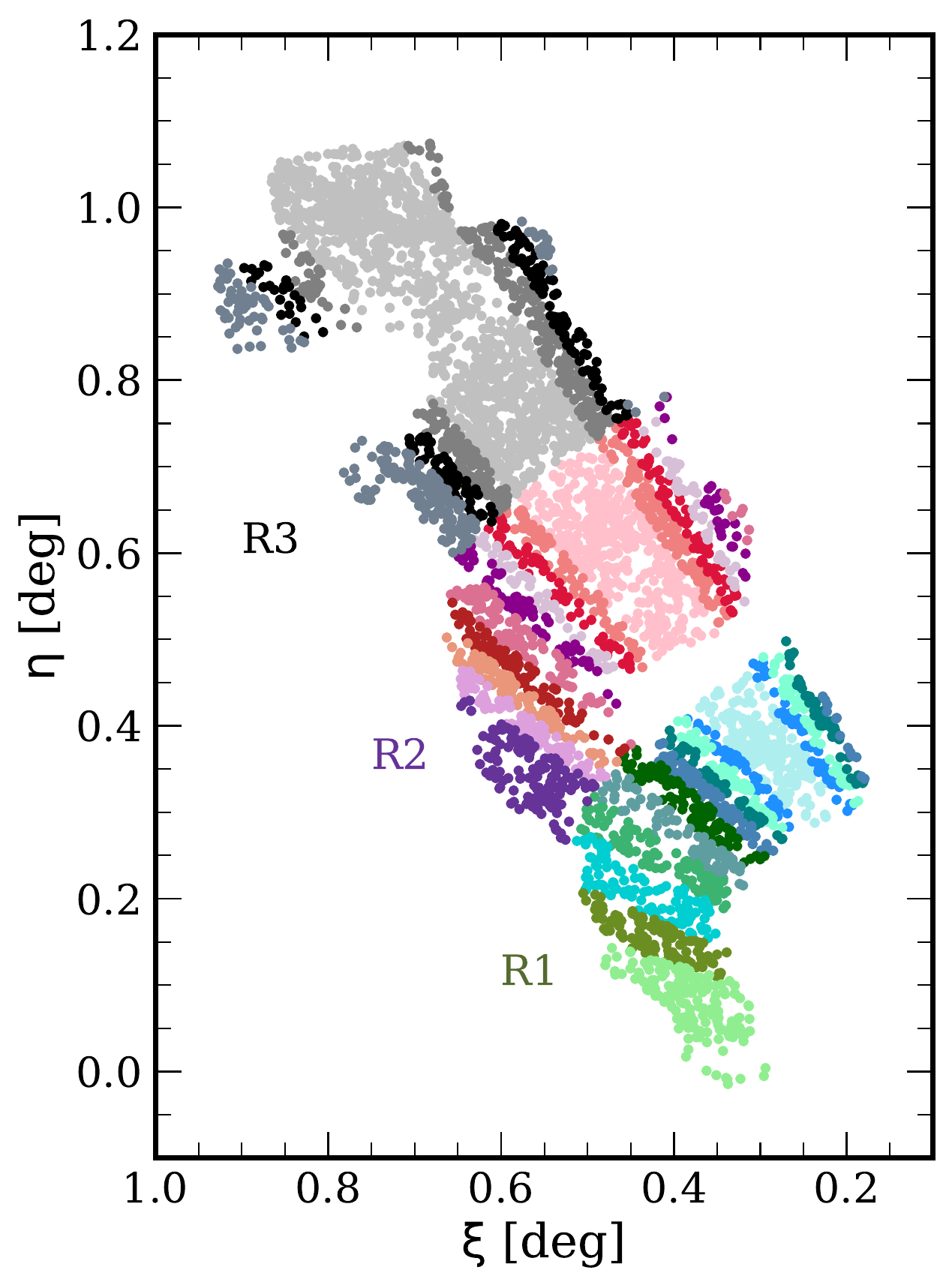}
    \caption{Location of annular radial regions (R1, R2, R3) and angular subregions in M31-centric coordinates used to model the velocity distributions (Section~\ref{sec:disk_regions}). 
    Each point corresponds to a M31 giant star with a succesful velocity measurement from SPLASH. 
    The regions R1, R2, and R3 are bounded by \rproj\ = 8, 12, and 18.5 kpc, where the kinematical parameters of the halo component in each region are fixed to the values obtained by \citet{Dorman2012}. Each radial annulus is divided into angular subregions based on absolute position angle relative to the major axis (P.A.\@ = 38$^\circ$), where these subregions are used to constrain the disk component of the velocity distribution. Note that stars on either side of the major axis represented by the same color are part of the same subregion. 
    Subregion designations are based on absolute angular distance from the major axis (i.e., R3$_1$ straddles the major axis, and R3$_4$ is the most distant from the major axis).} 
    \label{fig:disk_regions}
\end{figure}

\subsection{Membership} \label{sec:members}

We identified MW foreground stars based on the presence of the surface-gravity sensitive Na I $\lambda$8190 \AA\ doublet and on CMD position. The Na I doublet can discriminate between dwarf stars with strong Na I absorption features and giant stars with weak features \citep{Schiavon1997}, whereas the CMD can be used to select regions likely populated by giant stars. In total, we classified \nmembersplash\ stars as M31 members.

First, we assigned stars with foreground-extinction-corrected F475W$_0$--F814W$_0$ colors (Section~\ref{sec:zinit}) bluer than the most metal-poor \edit1{4} Gyr PARSEC RGB iscochrone \citep{Marigo2017}
by more than the photometric uncertainty to the MW foreground \edit1{(Section~\ref{sec:phot_params})}. These stars are substantially more likely to be intervening MW dwarfs \citep{Gilbert2006}, especially given that the SPLASH target selection procedure does not account for interstellar reddening (Section~\ref{sec:zinit},~\ref{sec:dust_map}) to favor a complete sample of giant star candidates. Alternatively, these blue stars may also be M31 disk stars in a different stellar evolutionary stage. SPLASH includes a large population of main-sequence turn-off stars in M31 (Figure~\ref{fig:members}), which is near the most contaminated portion of the CMD bounded by $m_{\rm F814W} < 21$ and $1 < m_{\rm F475W} - m_{\rm F814W} < 2$ \citep{Dorman2015}. 

Next, we evaluated the probability of M31 membership for stars based on their Na I equivalent widths (\ewna) and measurement uncertainties ($\delta$\ewna). We measured \ewna\ and $\delta$\ewna\ from all spectra following the method of \citet{Escala2020b}. \edit1{We were unable to measure \ewna\ for \percentnonasplash\ of SPLASH stars owing to factors such as weak absorption, convergence failure in line profile fits, or low S/N.\footnote{Note that weak absorption refers specifically to the lack of detectable absorption in at least one Na I line. This is distinct from a negative \ewna\ measurement (Figure~\ref{fig:members}), which can be caused by features such as poorly subtracted sky lines.} For stars with \ewna\ measurements,} we computed the membership probabilities from likelihood ratios derived by constructing non-parametric probability distribution functions (PDFs) using \ewna\ and $\delta$\ewna\ measurements (assuming Gaussian uncertainties) for thousands of MW and M31 stars securely identified in the SPLASH survey of M31's halo (e.g., \citealt{Gilbert2012}). The PDF weighted sample means and standard deviations are $\mu_{\rm Na, M31}$ = \meanewand\ \AA, $\sigma_{\rm Na, M31}$ = \stdewand\ \AA\ and $\mu_{\rm Na, MW}$ = \meanewmw\ \AA, $\sigma_{\rm Na, MW}$ = \stdewmw\ \AA. \edit1{For stars without \ewna\ measurements, we determined membership solely based on CMD position.\footnote{\fracgiantnoew\ (\fracrgbnowew) of M31 giant stars (RGB stars only) do not have \ewna\ measurements. In particular, \ewna\ measurements preferentially fail for stars above the TRGB. M31 giant stars (RGB stars only) without \ewna\ measurements are not biased in color compared to the full sample of M31 giant (RGB) stars.}} \edit1{Using the Besan\c{c}on model for the MW foreground \citep{Robin2003}, we expect that only $\sim$\fmwcontam\ of SPLASH stars in the M31 giant star region of the CMD are MW contaminants (Appendix~\ref{apdx:mw}).}

Figure~\ref{fig:members} shows the CMD of SPLASH disk stars color-coded by the \ewna-based M31 membership probability ($p_{\rm Na}$) calculated from the PDFs. Stars with $p_{\rm Na} \leq 0.25$, which are $\geq$3 times more likely to belong to the MW foreground than M31, preferentially populate the heavily contaminated portion of the CMD identified by \citet{Dorman2015}, in addition to the region above the TRGB, which has a higher incidence of contamination by bright MW foreground stars \citep{Gilbert2006}. 
Figure~\ref{fig:members} also demonstrates that stars with $p_{\rm Na} \leq 0.25$ tend to have MW-like velocities (\vhelio$_{,\rm MW}$ $\sim -50$ \kms), while few of these stars are present at velocities highly consistent with M31's stellar halo (\vhelio\ $< -300$ \kms). In contrast, defining MW foreground stars by $p_{\rm Na} < 0.5$, or $>$1 times more likely to belong to the MW than M31 based on \ewna\ alone, would exclude many targets with CMD positions and velocities fully consistent with M31 membership. The high surface density of M31's disk relative to the MW foreground ensures that the majority of targets are true M31 giant stars. We thus adopted $p_{\rm Na} \leq 0.25$ in addition to our CMD criterion to eliminate MW contaminants. \edit1{The CMD and Na I criterion classify \nremovecmd\ and \nremovena\ stars respectively as non-members.}

\begin{figure*}
    \centering
    \includegraphics[width=\textwidth]{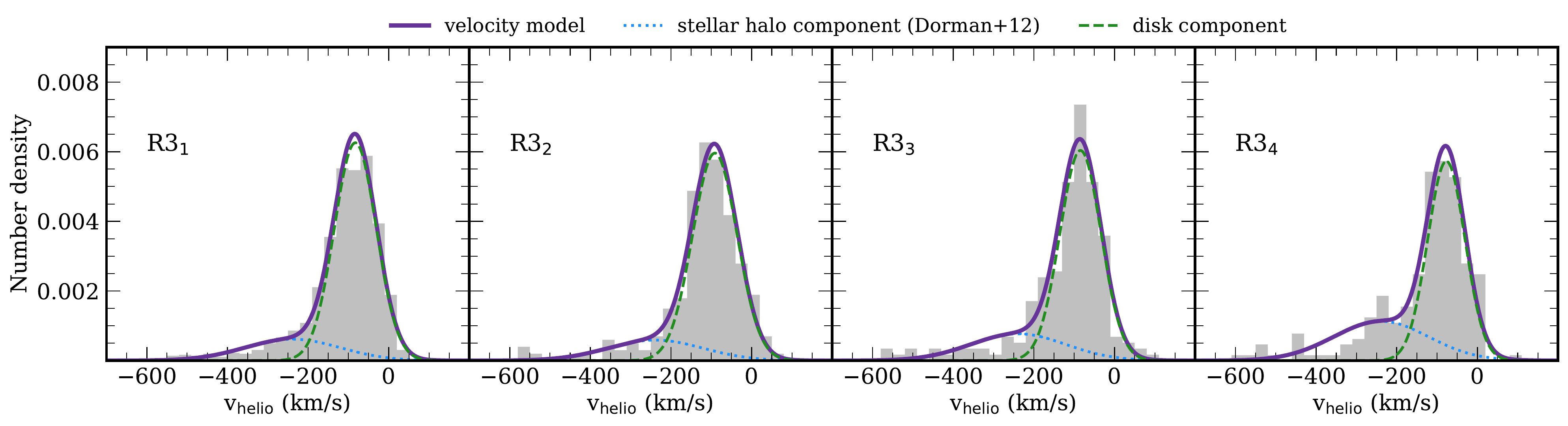}
    \caption{Heliocentric velocity distributions for region R3 (Section~\ref{sec:disk_regions}). From left to right, the panels show the velocity distribution in each subregion (grey histograms; Figure~\ref{fig:disk_regions}; Table~\ref{tab:vmodel}), where R3$_1$ straddles the major axis and R3$_4$ is the most distant from the major axis. We also show the fitted velocity model for each subregion (thick purple lines; Section~\ref{sec:disk_model}), which is composed of a stellar halo component \citep{Dorman2012} with fixed mean and dispersion but variable fractional contribution (dotted blue lines) and a disk component (dashed green lines). Figures~\ref{fig:ne1} and~\ref{fig:ne2} show the velocity distributions and models for R1 and R2. In general, the trends between the disk component velocity and $\Delta$P.A.\@ follow that expected for an inclined rotating disk, approaching M31's systemic velocity ($-300$ \kms) with increasing $\Delta$P.A. \citep{Dorman2012}. The stellar halo component also becomes more dominant with increasing $\Delta$P.A.\@
    }
    \label{fig:ne3}
\end{figure*}

This membership determination method stands in contrast to our previous work, which relied on \ewna, calcium-triplet based metallicity, CMD position, and radial velocity to probabilistically evaluate membership for stars along the line-of-sight to M31 (e.g., \citealt{Gilbert2006,Escala2020b}). The primary reason that we did not use these membership determination methods is 
to avoid relying on transformations between HST/ACS and Johnson-Cousins photometry (e.g., \citealt{Sirianni2005,Saha2011}) required for Ca II triplet based metallicity calibrations (e.g., \citealt{Ho2015}). Additionally, we did not use radial velocity as a diagnostic measurement for membership because the velocity distribution of M31's northeastern disk shows significant overlap with that of the MW's disk \citep{Ibata2005,Dorman2012}. Excluding radial velocity as a diagnostic does not preclude the use of the probablistic membership determination methods, but restricts their usage to two dimensions for stars without spectroscopic metallicity measurements, such that these methods do not confer a significant advantage over the adopted approach (see Section 3.4 of \citealt{Gilbert2006}). Again, the high stellar surface densities of M31's disk ensure that contamination from MW foreground stars is minimal prior to membership selection.



\subsection{Velocity Distribution Modeling} \label{sec:vmodel}

Previous studies suggest that M31's stellar velocity distribution in the disk region consists of a dynamically hot component and kinematically colder component(s) corresponding to the halo and disk respectively. To perform a chemodynamical analysis of M31's disk, we therefore separated stars likely belonging to the disk and halo over a spatial region spanning \rproj\ = 4--18.5 kpc.

\subsubsection{Disk Regions}
\label{sec:disk_regions}

\begin{table*}
    \centering
    \caption{Kinematical Model Parameters for the Disk}
    \begin{threeparttable}
    \begin{tabular*}{\textwidth}{@{\extracolsep{\fill}}lcccccccc}
        \hline\hline
        Subregion & \multicolumn{1}{p{1.0cm}}{\centering P.A.$_{+, i}$\\(deg)} & \multicolumn{1}{p{1.0cm}}{\centering P.A.$_{ +,f}$\\(deg)} & \multicolumn{1}{p{1.0cm}}{\centering P.A.$_{-, i}$\\(deg)} & \multicolumn{1}{p{1.0cm}}{\centering P.A.$_{ -,f}$\\(deg)} & \multicolumn{1}{p{1.0cm}}{\centering $\mu_{\rm disk}$\\(km/s)} & \multicolumn{1}{p{1.0cm}}{\centering $\sigma_{\rm disk}$\\(km/s)}& \multicolumn{1}{p{1.0cm}}{\centering $f_{\rm disk}$\\(km/s)} & $N_{\rm subregion}$\\ \hline
        \multicolumn{9}{c}{R1}\\ \hline
R1$_{1}$ & 38.0 & 41.9 & 34.1 & 38.0 &  $-$93.8$^{+6.3}_{-6.4}$ & 55.6$^{+4.7}_{-4.4}$ & 0.64$^{+0.05}_{-0.05}$ & 219\\
R1$_{2}$ & 41.9 & 43.8 & 32.2 & 34.1 &  $-$117.2$^{+7.8}_{-7.8}$ & 58.5$^{+5.0}_{-4.5}$ & 0.78$^{+0.06}_{-0.07}$ & 100\\
R1$_{3}$ & 43.8 & 45.7 & 30.3 & 32.2 &  $-$107.6$^{+9.6}_{-9.5}$ & 60.5$^{+6.4}_{-6.1}$ & 0.69$^{+0.07}_{-0.08}$ & 118\\
R1$_{4}$ & 45.7 & 47.6 & 28.4 & 30.3 &  $-$97.5$^{+13.5}_{-12.8}$ & 61.0$^{+7.4}_{-6.9}$ & 0.56$^{+0.09}_{-0.09}$ & 107\\
R1$_{5}$ & 47.6 & 49.8 & 26.2 & 28.4 &  $-$72.2$^{+10.7}_{-10.7}$ & 47.6$^{+5.5}_{-4.8}$ & 0.39$^{+0.07}_{-0.07}$ & 112\\
R1$_{6}$ & 49.8 & 53.4 & ... & ... &  $-$108.7$^{+9.6}_{-9.7}$ & 53.4$^{+5.7}_{-5.1}$ & 0.57$^{+0.08}_{-0.08}$ & 106\\
R1$_{7}$ & 53.4 & 57.5 & ... & ... &  $-$148.8$^{+10.3}_{-10.1}$ & 60.9$^{+7.4}_{-7.5}$ & 0.66$^{+0.09}_{-0.10}$ & 101\\
R1$_{8}$ & 57.5 & 61.7 & ... & ... &  $-$169.7$^{+6.9}_{-7.0}$ & 45.8$^{+5.5}_{-4.6}$ & 0.69$^{+0.08}_{-0.08}$ & 102\\
R1$_{9}$ & 61.7 & 67.7 & ... & ... &  $-$192.7$^{+7.5}_{-7.2}$ & 43.5$^{+5.7}_{-4.7}$ & 0.60$^{+0.08}_{-0.08}$ & 107\\
R1$_{10}$ & 67.7 & 72.9 & ... & ... &  $-$204.8$^{+6.8}_{-6.8}$ & 42.0$^{+4.7}_{-4.0}$ & 0.59$^{+0.08}_{-0.08}$ & 108\\
R1$_{11}$ & 72.9 & 92.7 & ... & ... &  $-$225.8$^{+7.5}_{-7.4}$ & 45.7$^{+4.7}_{-4.2}$ & 0.48$^{+0.07}_{-0.07}$ & 170 \\ \hline
        \multicolumn{9}{c}{R2}\\ \hline
R2$_{1}$ & 38.0 & 41.8 & 34.2 & 38.0 &  $-$78.9$^{+3.1}_{-3.1}$ & 54.3$^{+2.4}_{-2.2}$ & 0.88$^{+0.02}_{-0.02}$ & 421\\
R2$_{2}$ & 41.8 & 43.6 & 32.4 & 34.2 &  $-$89.5$^{+4.8}_{-4.8}$ & 60.6$^{+3.4}_{-3.2}$ & 0.90$^{+0.03}_{-0.03}$ & 215\\
R2$_{3}$ & 43.6 & 45.2 & 30.8 & 32.4 &  $-$98.9$^{+6.3}_{-6.3}$ & 59.3$^{+4.2}_{-4.0}$ & 0.86$^{+0.04}_{-0.05}$ & 140\\
R2$_{4}$ & 45.2 & 46.6 & 29.4 & 30.8 &  $-$91.6$^{+8.0}_{-8.2}$ & 65.6$^{+5.4}_{-5.0}$ & 0.84$^{+0.05}_{-0.06}$ & 112\\
R2$_{5}$ & 46.6 & 48.4 & 27.6 & 29.4 &  $-$90.3$^{+7.1}_{-7.2}$ & 51.6$^{+5.3}_{-4.7}$ & 0.77$^{+0.06}_{-0.06}$ & 104\\
R2$_{6}$ & 48.4 & 50.9 & 25.1 & 27.6 &  $-$99.2$^{+6.5}_{-6.4}$ & 53.0$^{+4.2}_{-3.9}$ & 0.83$^{+0.05}_{-0.05}$ & 102\\
R2$_{7}$ & 50.9 & 52.7 & ... & ... &  $-$124.0$^{+8.1}_{-8.5}$ & 55.1$^{+5.7}_{-5.1}$ & 0.75$^{+0.07}_{-0.07}$ & 101\\
R2$_{8}$ & 52.7 & 54.5 & ... & ... &  $-$106.8$^{+7.3}_{-7.2}$ & 53.3$^{+4.7}_{-4.2}$ & 0.71$^{+0.06}_{-0.06}$ & 103\\
R2$_{9}$ & 54.5 & 56.4 & ... & ... &  $-$139.4$^{+10.2}_{-10.6}$ & 54.9$^{+7.4}_{-6.1}$ & 0.59$^{+0.09}_{-0.09}$ & 103\\
R2$_{10}$ & 56.4 & 63.4 & ... & ... &  $-$142.6$^{+7.9}_{-7.8}$ & 59.9$^{+5.7}_{-5.2}$ & 0.68$^{+0.07}_{-0.07}$ & 146\\ \hline
        \multicolumn{9}{c}{R3}\\ \hline
R3$_{1}$ & 38.0 & 41.8 & 34.2 & 38.0 &  $-$83.0$^{+2.0}_{-2.0}$ & 52.2$^{+1.5}_{-1.4}$ & 0.82$^{+0.02}_{-0.02}$ & 1202\\
R3$_{2}$ & 41.8 & 43.6 & 32.4 & 34.2 &  $-$91.2$^{+3.8}_{-3.8}$ & 55.3$^{+2.8}_{-2.6}$ & 0.83$^{+0.03}_{-0.03}$ & 335\\
R3$_{3}$ & 43.6 & 45.1 & 30.9 & 32.4 &  $-$84.4$^{+5.1}_{-5.1}$ & 51.1$^{+3.8}_{-3.6}$ & 0.77$^{+0.04}_{-0.05}$ & 195\\
R3$_{4}$ & 45.1 & 49.7 & 26.3 & 30.9 &  $-$77.0$^{+4.6}_{-4.7}$ & 46.5$^{+3.4}_{-3.1}$ & 0.67$^{+0.04}_{-0.04}$ & 215\\
        \hline
    \end{tabular*}
    \begin{tablenotes}[flushleft]
    \item Note. \textemdash\ Disk regions are based on projected radial distance: R1, R2, and R3 are bounded by \rproj\ = 8, 12, and 18.5 kpc. Each region is divided into subregions based on absolute angular distance from the major axis (P.A. = 38$^\circ$; Section~\ref{sec:disk_regions}), where we 
    indicate the positive and negative P.A. boundaries (relative to the major axis) of each symmetric subregion in degrees east of north.
    The parameters describing the disk component in each subregion are velocity ($\mu$), velocity dispersion ($\sigma$), and normalized fractional contribution ($f$). The stellar halo component in each region is fixed to the values determined by \citet{Dorman2012} when correcting for tidal debris. The parameter values are the 50$^{\rm{th}}$ percentiles of the marginalized posterior probability distributions, where the errors are calculated from the 16$^{\rm{th}}$ and 84$^{\rm{th}}$ percentiles (Section~\ref{sec:disk_model}).
    \end{tablenotes}
    \end{threeparttable}
    \label{tab:vmodel}
\end{table*}

We followed the methodology of \citet{Dorman2012}, where we assumed only that the stellar disk is locally cold with a symmetric velocity distribution and that each region of M31's disk has a contribution from the inner halo. We similarly divided the disk into regions along the northeast major axis based on projected radial distance: R1, R2, and R3, which are bounded by \rproj\ = 8, 12, and 18.5 kpc. The median radius of stars in each bin is \medradneone, \medradnetwo, and \medradnethree\ kpc respectively. To model the velocity distribution, we fixed the halo component in each region using the Gaussian parameters determined by \citet{Dorman2012}, which are corrected for the presence of tidal debris at \vhelio\ $<$ $-500$ \kms\ likely related to the GSS \citep{Fardal2013,Escala2022}. The stellar halo component in regions (R1, R2, R3) is described by $\mu_{\rm halo}$ = (\muhaloone, \muhalotwo, \muhalothree) \kms\ and $\sigma_{\rm halo}$ = (\sighaloone, \sighalotwo, \sighalothree) \kms.

To fit for the disk component, we divided each region by position angle relative to the major axis ($\Delta$P.A.). As in \citet{Dorman2012}, we defined the angle subtended by each subregion as either the $\Delta$P.A.\@ that contains 100 stars at minimum, or the $\Delta$P.A.\@ such that the
predicted change in the line-of-sight velocity of M31's disk ($v_{\rm obs}$) owing to $\Delta$P.A.\@ is 10 \kms\ (comparable to the velocity measurement precision; Section~\ref{sec:splash}), choosing whichever was larger. As noted by \citet{Dorman2012}, the predicted velocity spread due to $\Delta$P.A.\@ within a subregion merely approximates the true spread, which is affected by additional factors such as the change in de-projected radius, deviations from perfectly circular rotation, and variations in the intrinsic local velocity distribution.

We calculated $v_{\rm obs}$ using a simple model for the perfectly circular rotation of an inclined disk \citep{Guhathakurta1988} with inclination angle $i=77^\circ$ and major axis P.A. = $38^\circ$ given by,
\begin{equation}
v_{\rm obs} (\xi, \eta) = v_{\rm sys} + \frac { v_{\rm rot} \sin (i) }{\sqrt{ 1 + \tan^2 (\Delta {\rm P.A.} ) / \cos^2 (i) }},
\end{equation}
where ($\xi$, $\eta$) are M31-centric tangent plane coordinates, $v_{\rm sys}$ = $-$300 \kms\ is the systemic velocity of M31, 
and $v_{\rm rot}$ = 250 \kms\ is the disk rotation speed,
which is approximately the median value measured from HI kinematics for de-projected radii in the disk plane ($R_{\rm disk} > 5$ kpc; \citealt{Chemin2009}). Figure~\ref{fig:disk_regions} shows the locations of the regions and subregions in M31-centric coordinates, where we divided R1, R2, and R3 into \edit1{11}, 10, and 4 subregions, respectively. Table~\ref{tab:vmodel} summarizes the spatial properties of each subregion bounded by lines of constant P.A., where we designate each subregion based on absolute angular distance from the major axis. 

\subsubsection{Fitting for the Disk Contribution}
\label{sec:disk_model}

We modeled the velocity distribution for each subregion using a two-component Gaussian mixture composed of a kinematically hot halo and colder disk \citep{Dorman2012}. The log likelihood function for a given subregion is described by,
\begin{equation}
    \ln \mathcal{L} = \sum_{i=1}^{N_{r,s}} \ln \left( f_s \mathcal{N} (v_i | \mu_s, \tau_s^{-1}) + (1 - f_s) \mathcal{N} (v_i | \mu_{r}, \tau_{r}^{-1} \right),
\label{eq:vmodel}
\end{equation}
where $i$ is an index corresponding to a M31 giant star with successful velocity measurement $v_i$ located in subregion $s$ within region $r$ and $N_{r,s}$ is the total number of such stars. Each Gaussian distribution $\mathcal{N}$ has mean velocity $\mu$ and inverse variance $\tau = 1/\sigma^2$, where $\mu_r$ and $\sigma_r$ are constant within a given region and correspond to the halo component. The fractional contribution of the disk, $f_s$, varies across subregions, where the halo fraction is constrained to $f_r = 1 - f_s$.

We sampled from the posterior probability distribution of Eq.~\ref{eq:vmodel} using an affine-invariant Markov chain  Monte  Carlo  (MCMC)  ensemble sampler ({\tt emcee}; \citealt{Foreman-Mackey2013}) with 10$^{2}$ walkers and $10^{4}$ steps. We implemented flat priors for $\mu_s$ and $f_s$ over the parameter ranges of [$-600$, $+100$] \kms\ and [0, 1], respectively. We assumed a Gamma prior on $\tau_s$ with $\alpha \approx 14$ and $\beta \approx 28,476$, which penalizes values of $\sigma_s$ below roughly 35 \kms\ and above 70 \kms\ based on the results of \citet{Dorman2012}, and boundary conditions on $\sigma_s$ of [5, 150] \kms. We determined the kinematical parameters for the disk component in each subregion from the latter 50\% of the samples, where Table~\ref{tab:vmodel} summarizes these parameters in terms of the 16$^{\rm th}$, 50$^{\rm th}$, and 84$^{\rm th}$ percentiles of the marginalized posterior distributions. Figure~\ref{fig:ne3} provides an example of the velocity distribution models fit to the four subregions in region R3. We show the fits to the subregions in R1 and R2 in Appendix~\ref{apdx:vmodel}. 

In general, the trends between the mean velocity of the disk component and absolute $\Delta$P.A. from the major axis follow that expected for an inclined rotating disk, in agreement with \citet{Dorman2012}, where the disk velocity approaches M31's systemic velocity ($-300$ \kms) with increasing angular distance. This trend is not apparent in R3 (Figure~\ref{fig:ne3}) because it spans a smaller angular range in position angle, but this relationship is clearly visible in Table~\ref{tab:vmodel} (see also Figures~\ref{fig:ne1} and~\ref{fig:ne2} in Appendix~\ref{apdx:vmodel}). We also recover that the halo component becomes more dominant with increasing angular distance from the major axis.

We emphasize that the purpose of the modeling is not to perform a detailed structural decomposition of the disk (e.g., into a thin and thick disk; Section~\ref{sec:thick_disk}), but rather to reliably distinguish disk stars from halo stars for the interpretation of the metallicity distributions (Section~\ref{sec:results}). As discussed by \citet{Dorman2012}, the assumption of a single locally cold disk component in each subregion provides a good fit to the velocity distributions and is sufficient for a simple decomposition. We computed a probability of belonging to the disk ($p_{\rm disk}$) for M31 giant stars based on their observed velocities and positions.
Using the velocity model for each star's assigned subregion (Table~\ref{tab:vmodel}), we calculated $p_{\rm disk}$ from likelihood ratio of the disk and halo components.

\begin{figure}
    \centering
    \includegraphics[width=\columnwidth]{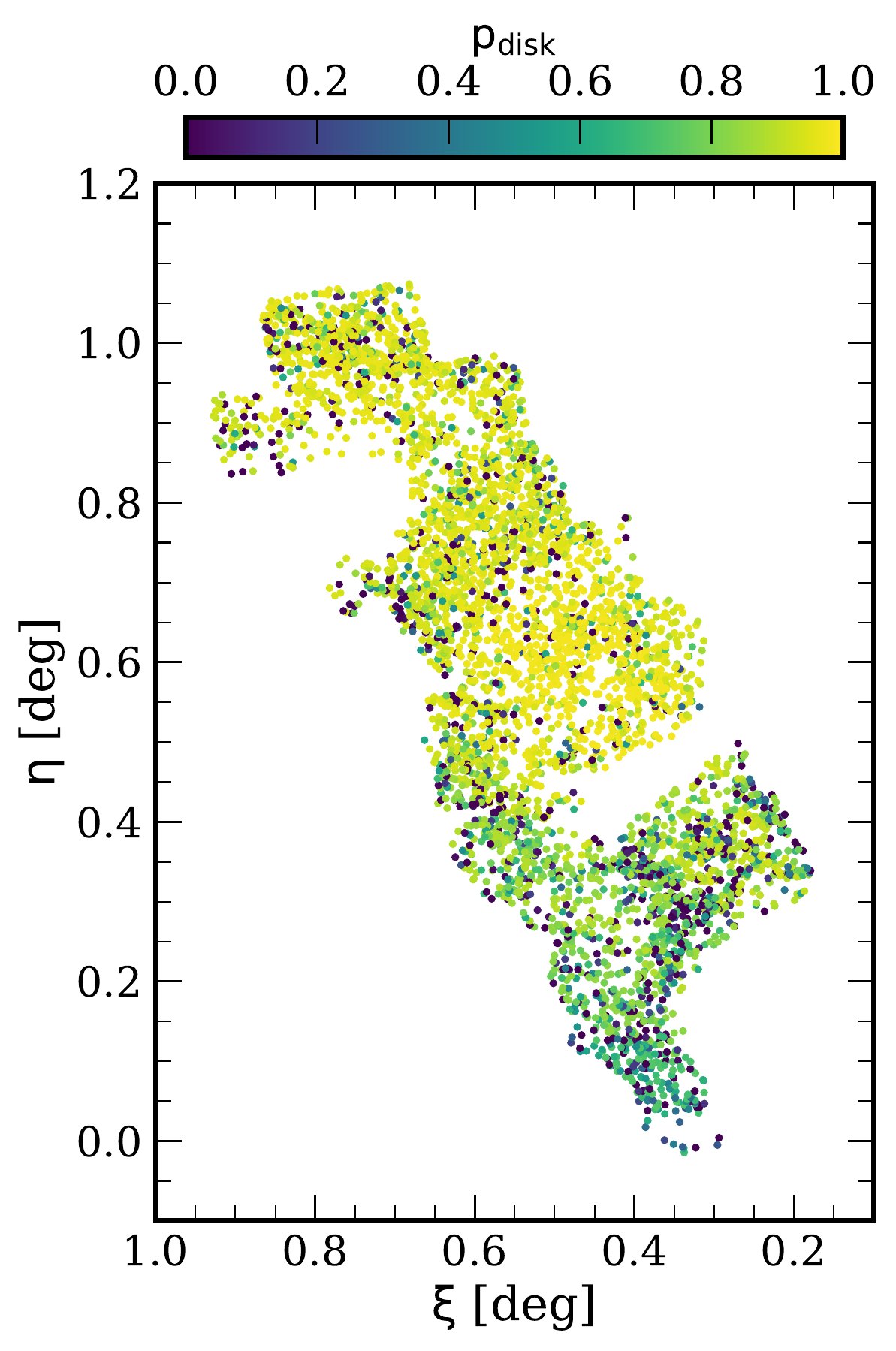}
    \caption{Location of giant stars in M31-centric coordinates, color-coded by the probability that a given star belongs to the disk versus the halo. The disk probability ($p_{\rm disk}$) is based on the velocity model for the subregion in which a star is located and its heliocentric velocity (Section~\ref{sec:disk_model}). Stars with $p_{\rm disk} \geq 0.75$ ($p_{\rm disk} \leq 0.25$) are $\geq$3 times more likely to belong to the disk versus the halo (and vice versa).
    }
    \label{fig:disk_prob}
\end{figure}

Figure~\ref{fig:disk_prob} illustrates the spatial distribution of M31 giant stars color-coded by the disk probability, where $p_{\rm disk}$ is highest near the major axis and in the outer disk.
Although the disk dominates star counts in the inner surveyed region \citep{Courteau2011,Dorman2013}, the halo's fractional contribution to the velocity distribution is also larger at small radii, resulting in lower values of $p_{\rm disk}$.
Over the SPLASH survey region, we found that \percentdisksplash\% (\percenthalosplash\%) of M31 giant stars have a high likelihood ($p_{\rm disk} >$ 0.75, or $p_{\rm disk} <$ 0.25 for the halo) of belonging to the disk (halo). The remaining giant stars (\percentmixedsplash\%) belong to a ``mixed'' population.

\section{Photometric Metallicity Measurement}
\label{sec:phot_params}

\begin{figure*}
    \centering
    \includegraphics[width=0.8\textwidth]{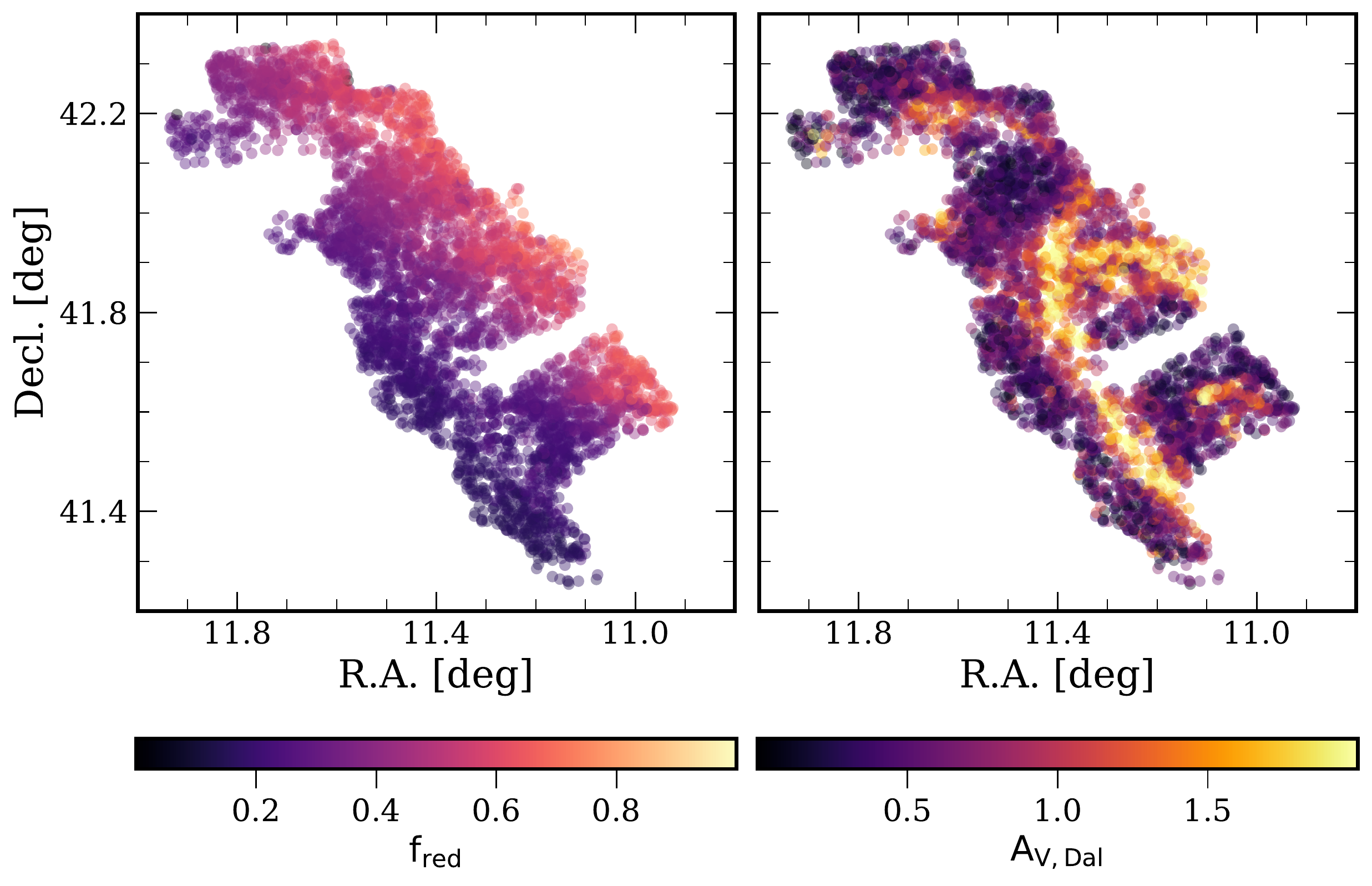}
    \caption{Sky location of M31 RGB stars in SPLASH color-coded by (left) the fraction of reddened RGB stars ($f_{\rm red}$) and (right) the median extinction ($A_{V, \rm Dal}$) from the maps of dust in M31's disk by \citet{Dalcanton2015} (Section~\ref{sec:dust_map}). We do not show the spread in extinction, where $\sigma_V \sim$ \sigmaavmed\ over the entire sample. RGB stars with $\Delta$P.A. $<$ 0 (Section~\ref{sec:disk_regions}) are more likely to be reddened given that M31's inclined thick disk is viewed in projection (J.~Dalcanton et al., in preparation). The dust extinction is highest in the 10 kpc star-forming ring \citep{Gordon2006}. We used these maps to assess the impact of dust in M31's disk on photometric metallicity measurements in Section~\ref{sec:fancy_dust}.
    }
    \label{fig:dust}
\end{figure*}

We measured photometric metallicity for M31 RGB stars accounting for all sources of dust extinction. We determine an initial metallicity by correcting for foreground reddening in Section~\ref{sec:zinit} and internal reddening due to M31's gaseous disk in Section~\ref{sec:zfinal}.

\subsection{Initial Metallicity Determination}
\label{sec:zinit}



We first corrected the PHAT photometry (Section~\ref{sec:phot}) for the effects of dust extinction caused by the MW foreground. We assumed $A_{V, {\rm MW}} = 0.2$ for the foreground reddening, which corresponds to the median value over the low extinction PAndAS footprint \citep{McConnachie2018} based on the dust maps by \citet{Schlegel1998} with corrections by \citet{SchlaflyFinkbeiner2011}. We used this value given that the foreground dust maps are inaccurate when restricted to M31's disk region where its dust emission dominates over that of the MW. This translates to A$_{\rm F814W} \sim 0.12$ and A$_{\rm F475W} \sim 0.38$ \citep{Gregersen2015}.


We classified \nrgbsplashmwcorr\  M31 members as red giants when 
correcting for foreground reddening. We defined the TRGB using 4 Gyr PARSEC isochrones \citep{Marigo2017} spanning $-2.2 <$ \fehphot\ $< +0.5$. We assumed a distance modulus of $m - M = 24.45 \pm 0.05$. Stars were assigned to the RGB if they are below the TRGB within the photometric uncertainty: $m_{\rm F814W} + \sigma_{\rm F814W} > m_{\rm TRGB}$ (median $\sigma_{\rm F814W}$ = \medrederr). The number of stars classified as red giants has a small dependence on the adopted foreground reddening, as well as the uneven reddening within M31's disk (Section~\ref{sec:dust_map}). 

The direction of the reddening vector may result in a few young AGB stars being reddened into the RGB CMD region, but old AGB stars will not be reddened into this region owing to the shape of the TRGB. The net effect of reddening is therefore to increase the number of stars classified as red giants. The predominant, but still minimal, source of contamination in the RGB region is red helium-burning stars 
given that MW dwarf stars can be distinguished from genuine giant stars (Section~\ref{sec:members}). Contamination by red helium-burning stars would increase the number of stars in the metal-poor tails of the predominately metal-rich distributions but should not bias their median values (Section~\ref{sec:mdfs}).


We determined the photometric metallicity for M31 RGB stars by interpolating de-reddened (F475W, F814W) photometry on a grid of 4 Gyr PARSEC isochrones in the relevant filters \citep{Escala2020a}. 
We did not extrapolate to determine \fehphot\ for stars blueward of the most metal-poor isochrone, thereby introducing an effective blue limit on the RGB region.

We assumed a relatively young age for RGB stars, although the mass-weighted average age for all stellar populations in M31's disk is 10 Gyr \citep{Williams2015,Williams2017}, because stars on the upper RGB are biased toward younger ages as a consequence of variable RGB lifetimes. \citet{Gregersen2015} used the star formation history measured over the PHAT footprint \citep{Williams2015} to simulate stellar populations in M31's disk, finding that it produced an upper RGB with a mean age of 4 Gyr. \citet{Dorman2015} found similar results for the mean RGB age when adopting a constant star formation history.
If we instead assume 10 Gyr ages for RGB stars, the median difference in the photometric metallicity is \difffehphotrgb\ when accounting for foreground reddening.
The assumed stellar age therefore affects the absolute metallicity scale, although we are primarily concerned with relative metallicities in this work. If the dominant halo population is systematically older than the assumed 4 Gyr old disk, the halo metallicity scale would decrease by a maximum of 0.26 (Section~\ref{sec:halo_comp}).

\subsection{Final Metallicity Determination}
\label{sec:zfinal}

\begin{figure}
    \centering
    \includegraphics[width=\columnwidth]{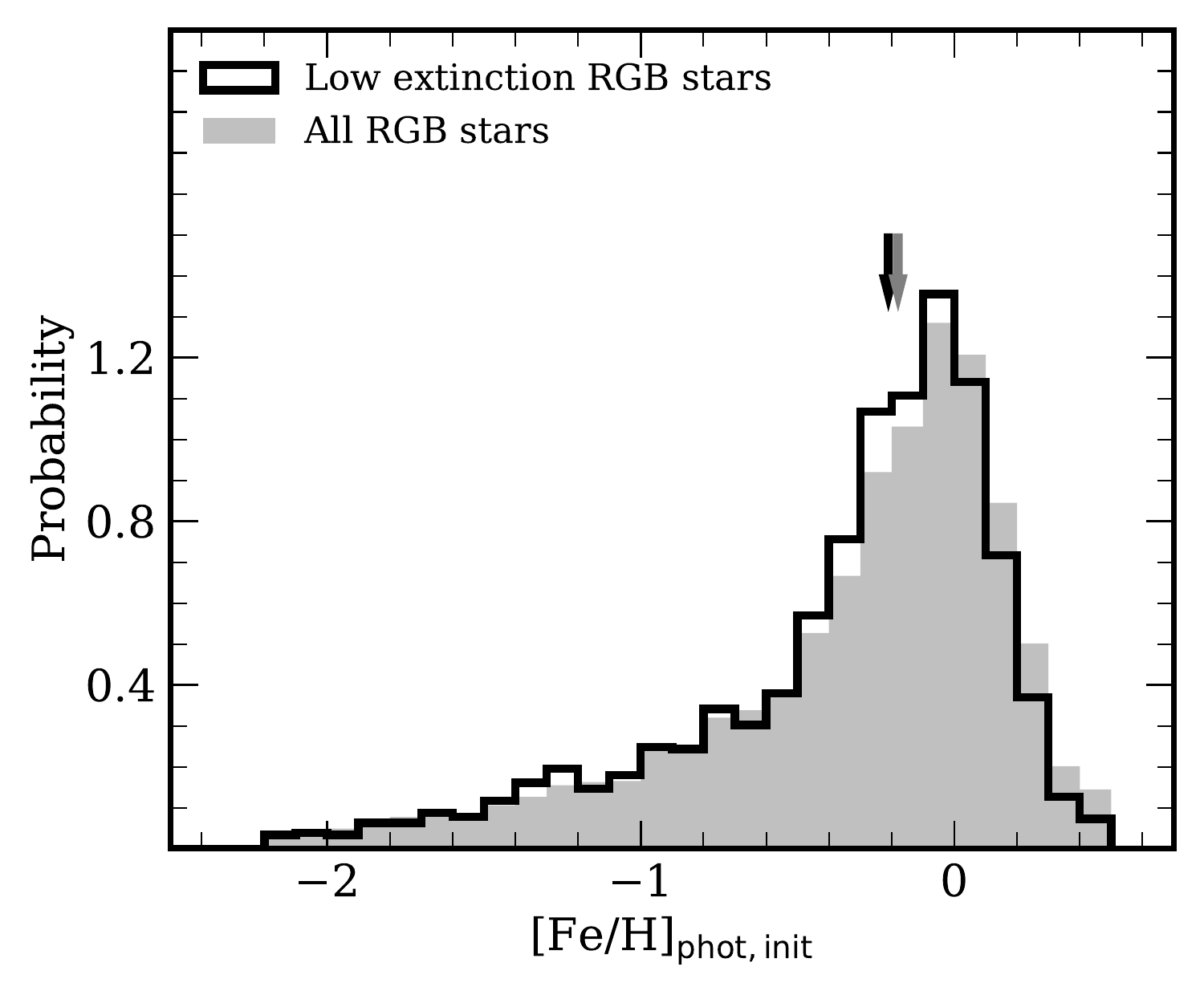}
    \caption{\fehphotinit\ distributions for all RGB stars (gray filled histogram) and RGB stars located in low extinction regions of the disk (black open histogram). We measured \fehphotinit\ using 4 Gyr PARSEC RGB isochrones on the foreground-reddening corrected optical CMD (Section~\ref{sec:phot_params}).
    We define low extinction regions using $f_{\rm red} \times A_{V, {\rm Dal}} < 0.25$ (\citealt{Gregersen2015}). The median \fehphotinit\ for all (low extinction) RGB stars is shown as a gray (black) arrow, where [Fe/H]$_{\rm phot, init, med}$ = \fehphotmedallrgb\ (\fehphotmedlowextrgb). This implies that most RGB stars in SPLASH are located in front of the dust layer and not significantly reddened. 
    }
    \label{fig:mdf_lowext}
\end{figure}

In this section, we modify our initial metallicity measurement (Section~\ref{sec:zinit}), which accounts only for foreground reddening, to correct for uneven reddening due to M31's disk. We adopt \fehphot\ (\fehphotinit) to refer to the final (inital) metallicities. We introduce the internal reddening maps and perform a preliminary assessment of dust effects on the metallicity distribution in Section~\ref{sec:dust_map}. We describe the adopted internal extinction correction and final metallicity measurements in Section~\ref{sec:fancy_dust}.

\subsubsection{Internal Reddening Maps}
\label{sec:dust_map}

 We used the maps by \citet{Dalcanton2015}, who used a novel approach to directly measure dust extinction from infrared (IR) PHAT photometry of RGB candidates assuming constant dust extinction from the MW foreground.  Based on the difference between the unreddened and reddened RGB sequences over the PHAT footprint, \citet{Dalcanton2015} modeled the spatial variation in reddening using the following parameters: median extinction (\avdal), dimensionless width of the log-normal extinction distribution (\sigav), and the fraction of reddened stars (\fred). This latter parameter reflects the geometry of the dust relative to the RGB stars, where 1$-$\fred\ percent of stars are assumed to be located in front of the thin dust layer of M31's disk.

Figure~\ref{fig:dust} presents maps of \fred\ and \avdal\ for SPLASH RGB stars obtained from 2D interpolation of each star's position on the \citeauthor{Dalcanton2015}\@ dust maps. We omit showing \sigav\ since it is relatively constant at \sigav\ $\sim$ 0.30 over the survey region. 
Figure~\ref{fig:dust} shows that the value of \fred\ increases from the southeast to northwest edge of the survey footprint because the RGB population is located in a thick ($h_z$ = \diskscaleheight\ kpc) and moderately inclined ($i=77^\circ$) disk viewed in projection (J.~Dalcanton et al., in preparation), where the expected value of \fred\ is 50\% at the major axis location (P.A. = 38$^\circ$). The value of \avdal\ is highest in the 10 kpc star forming ring \citep{Gordon2006} but variable across the survey footprint. 

\begin{figure*}
    \centering
    \includegraphics[width=0.8\textwidth]{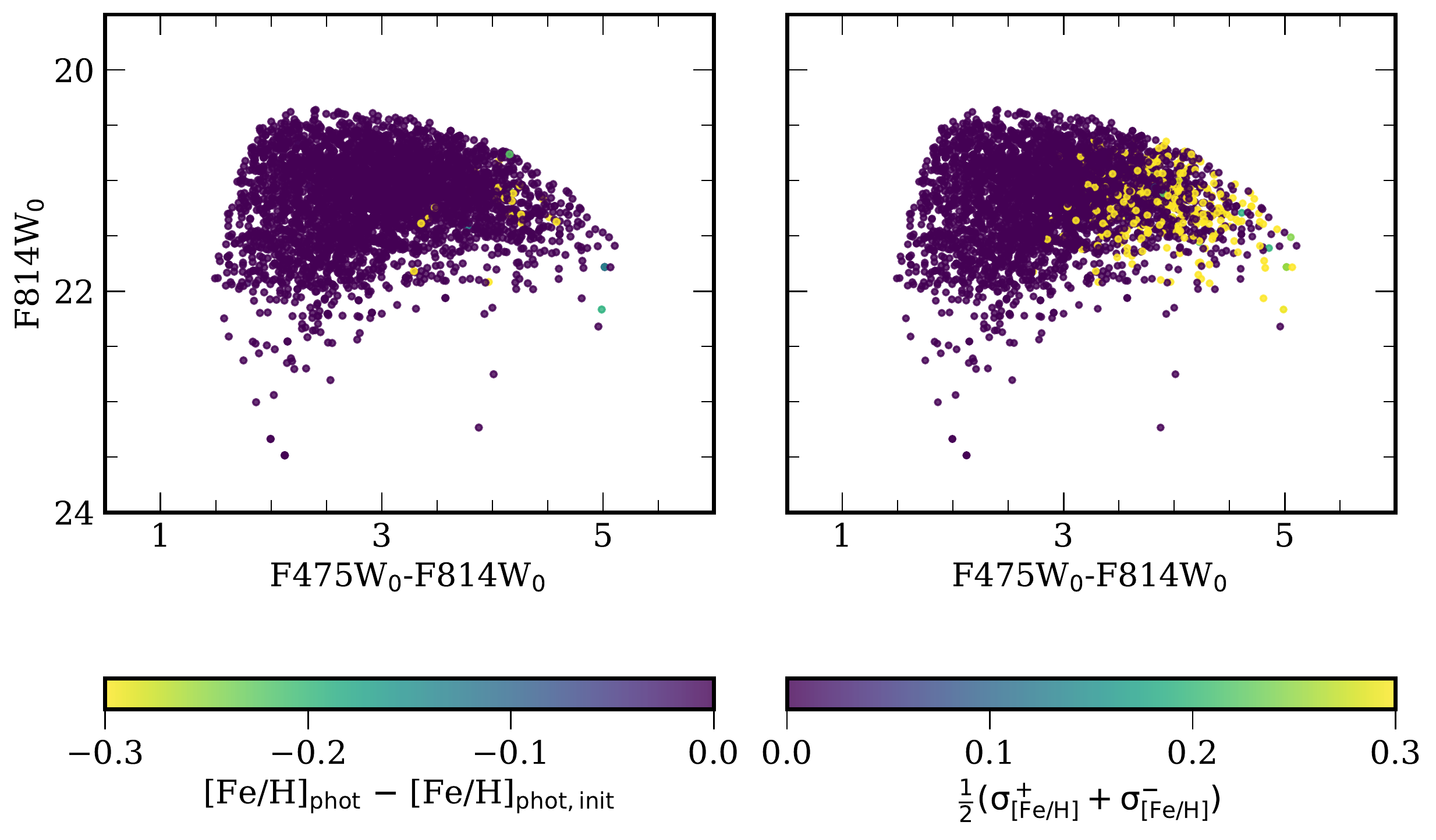}
    \caption{Foreground-reddening-corrected (F475W$_0$, F814W$_0$) CMDs for SPLASH RGB stars. 
    The CMDs are color-coded by (left) the difference between photometric metallicity computed with (\fehphot; Section~\ref{sec:zfinal}) and without (\fehphotinit; Section~\ref{sec:zinit}) taking into account reddening caused by dust in M31's disk and (right) the spread in \fehphot\ as quantified by the average of the 16$^{\rm th}$ ($\sigma_{\rm [Fe/H]}^-$) and 84$^{\rm th}$ ($\sigma_{\rm [Fe/H]}^+$) percentile errors of the \fehphot\ distribution. 
    We adopt \fehphot\ for stars with ($\sigma_{\rm [Fe/H]}^- + \sigma_{\rm [Fe/H]}^-)/2 < 0.03$ and exclude all other stars from the following analysis. We find that the median \fehphot\ $-$ \fehphotinit\  = \fehphotdiffrgbcorr\ 
     when accounting for both their CMD and spatial positions (cf.\@ Figure~\ref{fig:mdf_lowext}).
    }
    \label{fig:cmd_fehphotcorr}
\end{figure*}

One approach to account for the effects of dust in M31's disk is identifying RGB stars located in ``low-extinction'' regions. We defined low-extinction stars as those with \fred\ $\times$ \avdal\ $<$ 0.25 following \citet{Gregersen2015}, where \nrgblowext\ of \nrgbsplashmwcorr\ RGB stars meet this criterion when correcting for foreground reddening (Section~\ref{sec:zinit}). We then compared the metallicity distribution function (MDF) of RGB stars in low-extinction regions to the MDF of all RGB stars (Figure~\ref{fig:mdf_lowext}).
The median \fehphotinit\ for low-extinction RGB stars is [Fe/H]$_{\rm phot, init, med}$ = \fehphotmedlowextrgb\ compared to [Fe/H]$_{\rm phot, init, med}$ = \fehphotmedallrgb\ for all RGB stars. The low-extinction MDF differs most noticeably from the MDF for the full sample in the metal-rich peak, where the metal-poor tail is relatively unchanged.
These small metallicity differences 
indicate that the majority of SPLASH RGB stars are located in front of the dust layer and not significantly reddened (see Appendix~\ref{apdx:dust} for additional support from the IR CMDs). However, this approach is limiting given that it only considers \textit{spatial} information to assess whether a star is reddened, when unreddened RGB sequences are present in the IR CMD at each location (Section~\ref{sec:fancy_dust}).

\subsubsection{CMD-Based Extinction Correction}
\label{sec:fancy_dust}

We incorporated information on the optical CMD position, as opposed to solely using spatial position as in the case of the low-extinction regions (Section~\ref{sec:dust_map}), to account for the effect of dust in M31's disk on the metallicity determination. We 
constructed an extinction probability distribution for each RGB star with index $i$,
\begin{equation}
\label{eq:dust_pdf}
\begin{split}
    P_i(A_V|\alpha_i, \delta_i) &= C(f_{{\rm red},i}) \times \\ & \frac{1}{A_V \sqrt{2 \pi \sigma_{V,i} }} \exp \Big[ {- \frac{(\ln (A_V / A_{V, {\rm Dal}, i} ) )^2}{2 \sigma_{V,i}^2}} \Big]
\end{split}
\end{equation}
where $A_V$ is a variable representing $V$-band extinction, $A_{V, {\rm Dal},i}$, $\sigma_{V,i}$, and $f_{{\rm red},i}$ are the star's dust model parameters assigned from its sky coordinates ($\alpha_i, \delta_i$), and $C \in \{0,1\}$ is a constant sampled with probability \{1$-f_{{\rm red},i}$, $f_{{\rm red},i}$\}. We drew 10$^{3}$ values of $A_V$ from Eq.~\ref{eq:dust_pdf} for each star, which we then converted to $A_{\rm F475W}$ and $A_{\rm F814W}$ (Section~\ref{sec:zinit}). These HST-band extinction distributions were then used with the constant $A_{V, {\rm MW}}$ to shift each star's observed CMD position, thereby creating a statistical distribution of CMD positions corrected for all sources of dust extinction. We measured the metallicity (Section~\ref{sec:zinit}) for each corrected CMD position for each star given its fixed classification on the RGB (Section~\ref{sec:zinit}). 
From this distribution, we calculated a median \fehphot\ value corrected for all sources of dust extinction for each star.

Figure~\ref{fig:cmd_fehphotcorr} shows CMDs for RGB stars color-coded by the difference between \fehphot\ and \fehphotinit. It also shows the spread in \fehphot\ as quantified by the average of the 16$^{\rm th}$ ($\sigma_{\rm [Fe/H]}^-$) and 84$^{\rm th}$ ($\sigma_{\rm [Fe/H]}^+$) percentile errors of the \fehphot\ distribution. For most stars, \fehphot\ is unchanged from the original \fehphotinit\ value. 
We adopted \fehphot\ as the photometric metallicity for stars with precise \fehphot\ determinations (($\sigma_{\rm [Fe/H]}^- + \sigma_{\rm [Fe/H]}^+)/2 < 0.03$), where \fehphoterrmed\ is the median statistical metallicity uncertainty (\fehphoterr) from the propagation of photometric uncertainties.  
We also incorporated the spread in \fehphot\ as an error term contributing to the total metallicity uncertainty. Stars without precise \fehphot\ measurements are excluded from the following analysis.

\begin{figure*}
    \centering
    \includegraphics[width=\textwidth]{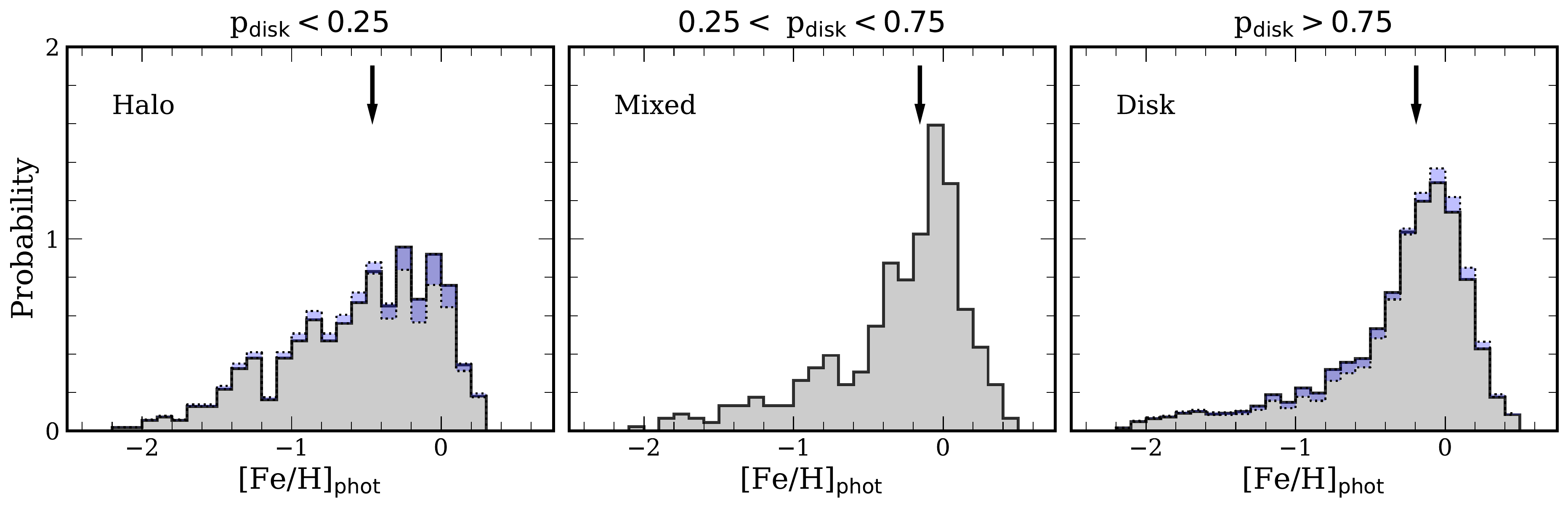}
    \caption{Metallicity distribution functions of RGB stars (black outlined gray filled histograms; Section~\ref{sec:mdfs}) corrected for all sources of dust extinction (\fehphot; Section~\ref{sec:zfinal}). We separate stars into halo (left; \pdisk\ $<$ 0.25; Section~\ref{sec:disk_model}), mixed (middle; $0.25 <$ \pdisk\ $< 0.75$), and disk (right; \pdisk\ $>$ 0.75) populations. 
    The bin size is 0.10, where the median $\delta$\fehphot\ is 0.03.
    The median metallicity for each population is indicated as an arrow. For the halo, mixed, and disk populations, the median \fehphot\ is \fehphotmedhalorgb, \fehphotmedmixedrgb, and \fehphotmeddiskrgb\ respectively (Table~\ref{tab:fehphot}). 
    The dotted histogram is the resulting halo (disk) MDF in the most extreme case of contamination by metal-rich disk (metal-poor halo) interlopers (Appendix~\ref{apdx:contam}). The blue shading defines the uncertainty region between the fiducial (solid) and contamination-corrected (dotted) MDFs.
    The MDFs for the disk and mixed populations are similar. The halo population is more metal-poor but still has metal-rich stars (Section~\ref{sec:mdfs}).
    }
    \label{fig:mdfs}
\end{figure*}

These selection criteria reduce the RGB 
sample from \nrgbsplashmwcorr\
to \nrgbdustcorr\ 
stars. The median difference between the final \fehphot\ and original \fehphotinit\ distribution is \fehphotdiffrgbcorr. 
We note that aside from the decrease in metallicity, the overall structure of the MDFs is unaltered, especially in the metal-poor regime. The metallicity difference 
is similar to that of the low extinction region selection (Figure~\ref{fig:mdf_lowext}). However, the approach involving Eq.~\ref{eq:dust_pdf} has the advantage of using both CMD and spatial information to construct a sample of RGB stars with relatively certain metallicity determinations despite the effects of dust in M31's disk. 

\section{Chemodynamics of the Disk and Halo}
\label{sec:results}

In this section, we analyzed the kinematical and chemical properties of RGB stars along the line-of-sight to M31's disk, separating them into disk, halo, and ``mixed'' subpopulations (Section~\ref{sec:vmodel}), to ultimately investigate evolutionary scenarios for the disk and halo. We present metallicity distribution functions (MDFs) and asymmetric drift measurements for each subpopulation in Sections~\ref{sec:mdfs} and~\ref{sec:ad} respectively. We measure metallicity gradients for the disk and halo in Section~\ref{sec:grad}.

\subsection{Metallicity Distribution Functions}
\label{sec:mdfs}

\begin{table}
\centering
\begin{threeparttable}
    \caption{Photometric Metallicity Distribution Properties for RGB Stars in the Halo, Mixed, and Disk Populations}
    \begin{tabular*}{\columnwidth}{lccc}
    \hline \hline
    Pop. & med\{\fehphot\} & $\langle$\fehphot$\rangle$ & $\sigma$\{\fehphot\} \\ \hline
    Halo &  $-0.46$ & $-0.55 \pm 0.02$ & 0.50 \\
    Mixed & $-0.16$ & $-0.31 \pm 0.02$ & 0.48 \\
    Disk & $-0.19$ & $-0.33 \pm 0.01$ & 0.51 \\
    \hline
    \end{tabular*}
    \begin{tablenotes}[flushleft]
    \item Note.\textemdash\ The columns are population and the simple median, mean, and standard deviation of its metallicity distribution, where \fehphot\ has been corrected for all sources of dust extinction (Section~\ref{sec:zfinal}). The populations are defined using the spatially and kinematically based probability that an RGB star belongs to the disk (Section~\ref{sec:disk_model},~\ref{sec:mdfs}).
    \end{tablenotes}
    \label{tab:fehphot}
\end{threeparttable}
\end{table}

\begin{figure*}
    \centering
    \includegraphics[width=\textwidth]{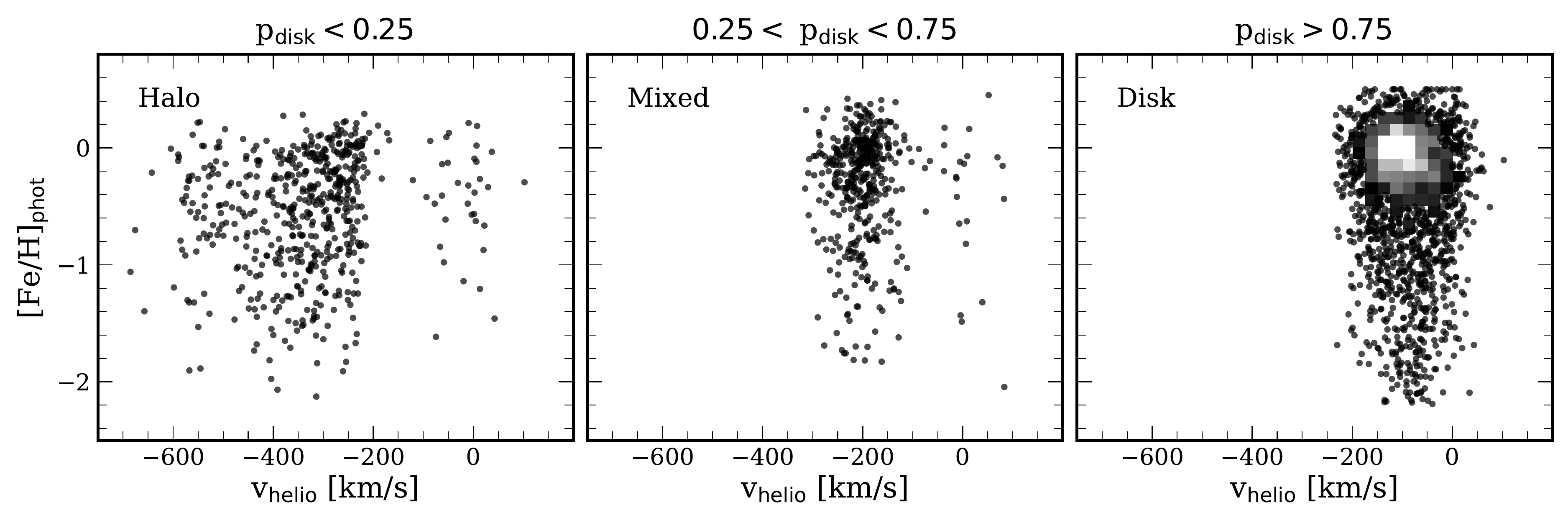}
    \caption{Photometric metallicity versus heliocentric velocity for RGB stars 
    in SPLASH (Section~\ref{sec:mdfs}). We omit showing \fehphot\ and \vhelio\ measurements uncertainties for clarity (typical values 0.03 and 5-10 \kms, respectively).
    From left to right, stars are separated into halo, mixed, and disk populations. We show a histogram in the higher density region of the disk population (bins 25 \kms\ and 0.1). The halo population is concentrated at \vhelio\ $\lesssim -200$ \kms\ owing to the high likelihood that stars with $-200$ \kms\ $\lesssim$ \vhelio\ $\lesssim$ 0 \kms\ belong to the dominant disk component (Table~\ref{tab:vmodel}). 
    Most disk stars are concentrated around \fehphot\ $\sim$ $-0.10$ and \vhelio\ $\sim$ $-100$ \kms, although a metal-poor tail is also present. The mixed population is dominated by stars with disk-like \fehphot\ but lower \vhelio\ $\sim$ $-200$ \kms. The halo shows evidence of a population with disk-like \fehphot\ at \vhelio\ $\sim$ $-280$ \kms\ and another that is kinematically hotter and more metal-poor.
    }
    \label{fig:vhelio_vs_fehphot}
\end{figure*}

We explored the MDFs of RGB stars corrected for all sources of dust extinction (Section~\ref{sec:zfinal}) in stellar halo and disk populations (Section~\ref{sec:vmodel}) along the line-of-sight to M31's disk. We defined disk (halo) stars using \pdisk\ $>$ 0.75 (\pdisk\ $<$ 0.25), which corresponds to stars that are at least 3 times more likely to belong to the disk (halo) based on kinematics and spatial position. We designated all other RGB stars as ``mixed.'' In total, \nhalorgb, \nmixedrgb, and \ndiskrgb\ stars belong to the halo, mixed, and disk populations respectively. Figure~\ref{fig:mdfs} shows the MDFs for each population and Table~\ref{tab:fehphot} summarizes their properties.

The disk MDF (median \fehphot\ = \fehphotmeddiskrgb) has a dominant metal-rich population and an extended metal-poor tail. Modeling the MDF using a two-component Gaussian mixture (analogous to Section~\ref{sec:disk_model}) yields \fehphot\ = \fehphotdiskrgbmr\ (\fehphotdiskrgbmp), $\sigma_{\rm [Fe/H]_{\rm phot}}$ = \fehphotdiskrgbmrsig\ (\fehphotdiskrgbmpsig) with a fractional contribution of \fehphotdiskrgbmrfrac\ (\fehphotdiskrgbmpfrac) for the metal-rich (metal-poor) disk population. The similarity between the mixed (median \fehphot\ = \fehphotmedmixedrgb) and disk MDFs implies that this intermediate population may be dominated by genuine disk stars located in subregions with larger halo fractions
\edit1{or similar mean velocities for the halo and disk} (Section~\ref{sec:thick_disk}), resulting in an uncertain separation between structural components. 
The halo MDF (median \fehphot\ = \fehphotmedhalorgb) is best described by a metal-rich (metal-poor) component with \fehphot\ = \fehphothalorgbmr\ (\fehphothalorgbmp), $\sigma_{\rm [Fe/H]_{\rm phot}}$ = \fehphothalorgbmrsig\ (\fehphothalorgbmpsig), and fractional contribution \fehphothalorgbmrfrac\ (\fehphothalorgbmpfrac). 

The presence of halo stars with similar metallicity to the disk is unusual given expectations that halos are typically metal-poor. However, the metal-rich halo population appears to be genuine, where $\lesssim$6\% of metal-rich halo stars are expected to be disk interlopers. In Appendix~\ref{apdx:contam}, we estimated this fraction by calculating the expected number of disk (halo) contaminants given the \pdisk-based definition for the halo (disk) population from the velocity models (Section~\ref{sec:vmodel}). We then assessed the maximal impact expected on the fiducial halo (disk) MDF from contamination by metal-rich disk (metal-poor halo) stars using the MDF models (Figure~\ref{fig:mdfs}).

\begin{figure*}
    \centering
    \includegraphics[width=\textwidth]{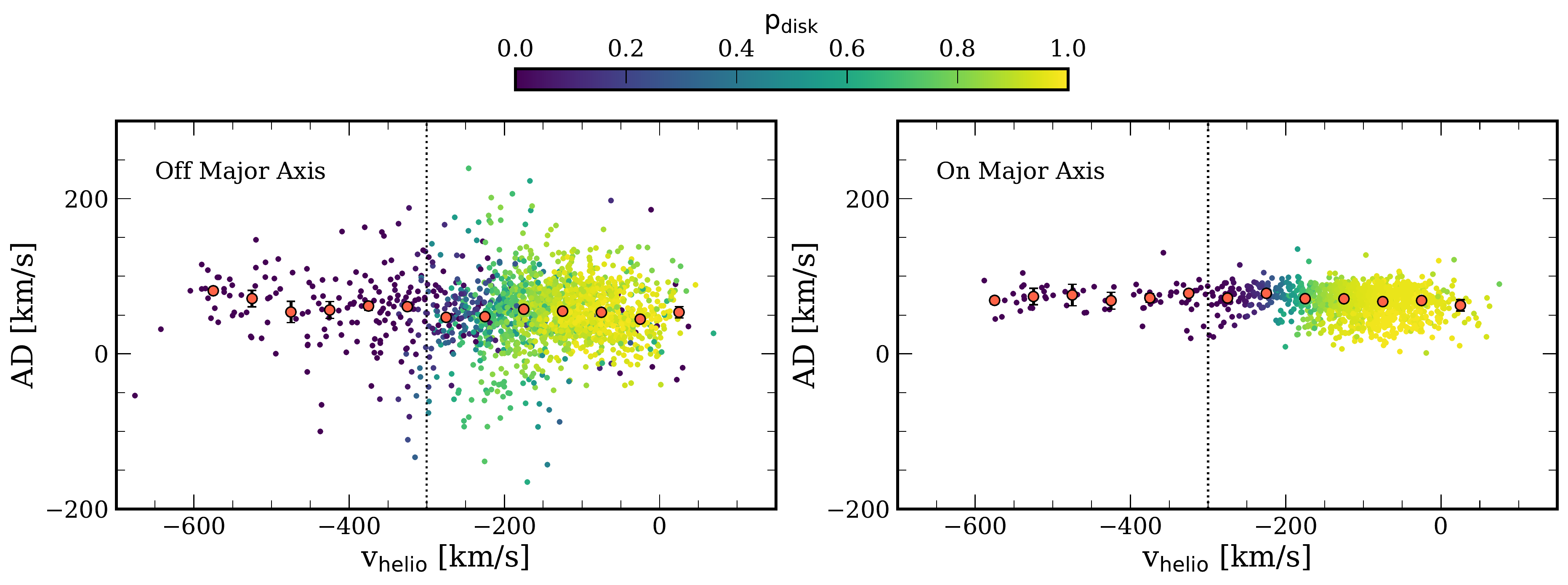}
    \caption{Asymmetric drift (AD) with respect to HI \citep{Quirk2019} versus heliocentric velocity for SPLASH RGB stars (Section~\ref{sec:ad}). We show stars located off (left) and on (right) the major axis, where the AD for off-axis stars exhibits more scatter due to geometrical effects. Each star is color-coded by its disk probability. The dotted line is M31's systemic velocity. The orange points are the median AD in 50 \kms\ \vhelio\ bins. The variation in AD is much larger between different stellar types than within the RGB population \citep{Quirk2019}. Despite this, a clear increase in AD is visible between 
    the disk (\pdisk\ $>$ 0.75) and halo (\pdisk\ $<$ 0.25) RGB populations. The  metal-rich group of halo stars (\vhelio\ $\sim$ $-280$ \kms, \fehphot\ $\sim$ $-0.10$; Figure~\ref{fig:vhelio_vs_fehphot}) lags the gaseous disk to a similar extent as the rest of the halo population.
    }
    \label{fig:ad_vs_vhelio}
\end{figure*}

Figure~\ref{fig:vhelio_vs_fehphot} shows the relationship between \fehphot\ and \vhelio\ for each population. The halo population is concentrated at \vhelio\ $\lesssim -200$ \kms\ owing to the high likelihood that stars with $-200$ \kms\ $\lesssim$ \vhelio\ $\lesssim$ 0 \kms\ belong to the disk (Table~\ref{tab:vmodel}) regardless of metallicity. Figure~\ref{fig:vhelio_vs_fehphot} further demonstrates that the dominant metal-rich disk population (\fehphot\ $\sim$ $-0.10$ and \vhelio\ $\sim$ $-100$ \kms) appears to continuously extend towards M31's systemic velocity, where a similarly metal-rich group of stars is evident at \vhelio\ $\sim$ $-200$ \kms\ ($-280$ \kms) in the mixed (halo) population. The metal-rich halo stars mainly belong to this group, where a second kinematically hotter population encompasses the majority of the metal-poor halo stars. In order to identify stars in the metal-rich group, we modeled the halo in metallicity versus velocity space as a combination of bivariate normal distributions,
\begin{multline}
    \mathcal{G}(v, x | \vec{\mu}, \vec{\sigma} ) = \frac{1}{2 \pi \sigma_{v} \sigma_{x} \sqrt{1 - r^2} } \\\times \exp \left( -\frac{1}{2 (1 - r^2)} \left[ \frac{ (v - \mu_v ) }{\sigma_{v}^2} + \frac{ (x - \mu_x ) }{\sigma_{x}^2} \right. \right. \\ \left. \left. - \frac{2 r (v - \mu_v) (x - \mu_x)}{\sigma_{v} \sigma_{x}} \right] \right),
    \label{eq:bi_normal}
\end{multline}
\begin{equation}
    \mathcal{G} = f_{\rm mp} \mathcal{G}_{\rm mp} + (1 - f_{\rm mp}) \mathcal{G}_{\rm mr},
\label{eq:2d_halo_model}
\end{equation}
where $v$ is \vhelio, $x$ is \fehphot, $\vec{\mu} = (\mu_v, \mu_x)$, $\vec{\sigma} = (\sigma_v, \sigma_x)$, and $r$ are the means, standard deviations, and correlation coefficient of the normal distribution. The halo is separated into a metal-poor population with fractional contribution $f_{\rm mp}$ and a metal-rich group with $f_{\rm mr} = 1 - f_{\rm mp}$. We sampled from the posterior probability distribution \citep{Foreman-Mackey2013}\footnote{We used 10$^2$ walkers, 10$^4$ steps, for a total of $5 \times 10^{5}$ samples from the latter 50\% of each chain. We assumed inverse Gamma priors on dispersion parameters and flat priors otherwise over a broad but reasonable set of parameter values. We required that $\mu_{x}^{\rm mp} < \mu_{x}^{\rm mr}$. The final parameters are computed from the 16$^{\rm th}$, 50$^{\rm th}$, and 84$^{\rm th}$ percentiles of the posterior probability distributions.
}
of Eq.~\ref{eq:2d_halo_model} to obtain $\vec{\mu}$ = (\muvhalomr\ $\pm$ \muverrhalomr\ \kms, \muxhalomr\ $\pm$ \muxerrhalomr), $\vec{\sigma}$ = (\sigvhalomr\ $\pm$ \sigverrhalomr\ \kms, \sigxhalomr\ $\pm$ \sigxerrhalomr), $r$ = \rhalomr\ $\pm$ \rerrhalomr, and $f_{\rm mr}$ = \fmrhalo\ $\pm$ \fmrerrhalo\ for the metal-rich group of halo stars. For the metal-poor group, we found $\vec{\mu}$ = (\muvhalomp\ $\pm$ \muverrhalomp\ \kms, \muxhalomp\ $\pm$ \muxerrhalomp), $\vec{\sigma}$ = (\sigvhalomp\ $\pm$ \sigverrhalomp\ \kms, \sigxhalomp\ $\pm$ \sigxerrhalomp), and $r$ = \rhalomp\ $\pm$ \rerrhalomp.

For each halo star, we calculated a  probability of belonging to the metal-rich group, $p_{\rm mr}$, based on the likelihood ratio between $\mathcal{G_{\rm mr}}$ and $\mathcal{G_{\rm mp}}$. We assigned stars to this group if $p_{\rm mr} > 0.75$, where $<$\fracpkickcontam\ are expected to be disk interlopers (Appendix~\ref{apdx:contam}). We further explore the properties of this interestingly metal-rich halo group---and the broader halo population---with respect to the rotation of M31's gaseous disk in Section~\ref{sec:ad}.

\subsection{Asymmetric Drift}
\label{sec:ad}

We examined the asymmetric drift (AD), or the difference between the gas and stellar rotation velocity at a given deprojected radius, for the various RGB subpopulations. We used AD measurements from \citet{Quirk2019} determined using SPLASH stellar velocities \citep{Dorman2012,Dorman2013,Dorman2015} and HI 21 cm data \citep{Chemin2009} in combination with a tilted ring model to derive stellar and gaseous rotation curves for M31's disk. \citeauthor{Quirk2019}\@ found that AD increases with mean stellar age as traced by various stellar types, where old RGB stars (4 Gyr) lag the gas the farthest at 63.0 \kms. The variation in AD between stellar types is significantly larger than within the population of a given stellar type such as RGB stars \citep{Quirk2019}.\footnote{We also note that AD measurements are only available for RGB stars with \fehphot\ $\gtrsim$ $-1$ due to the shape of the CMD-based RGB selection box used by  \citet{Quirk2019}. 
For the \percentmrhalo, \percentmrmixed, \percentmrdisk\ of stars in the halo, mixed, and disk populations that have \feh\ $>$ $-1$ and AD measurements, we find that AD is independent of metallicity for each RGB subpopulation. We therefore do not expect this metallicity bias to significantly impact the comparison of the relative AD between subpopulations.
}

Figure~\ref{fig:ad_vs_vhelio} shows the relationship between AD and \vhelio\ for RGB stars as defined in this work (Section~\ref{sec:members},~\ref{sec:zinit}) color-coded by disk probability. We separated stars into groups on and off the major axis, where the former contains stars in subregions that straddle the major axis (i.e., with the subscript ``1'' ; Table~\ref{tab:vmodel}) and the latter contains all other stars. The AD distribution for off-axis stars exhibits more scatter than those on-axis due to geometrical effects associated with measuring AD in an inclined disk \citep{Quirk2019}. Considering both on- and off-axis RGB stars,  the median AD is \admeddiskall, \admedmixedall, and \admedhaloall\ \kms\ for the disk, mixed, and halo populations, respectively. At face value, this suggests that the halo and disk ADs are marginally consistent at the \addiskhalodiffsigall$\sigma$ level and that the mixed and disk populations have fully consistent ADs. 

However, given the geometrical effects impacting the empirical AD measurements in areas off the major axis, the on-axis RGB stars provide a more precise representation of the underlying AD distribution.
In this case, the median AD increases between the disk (\admeddisk\ \kms) and the mixed (\admedmixed\ \kms) and halo (\admedhalo\ \kms) populations, in accordance with expectations of dynamically colder to hotter populations. The disk and halo ADs are distinct at the \addiskhalodiffsig$\sigma$ level, whereas the mixed population AD is consistent with the halo but differs from the disk by \addiskmixeddiffsig$\sigma$. The finding that the halo lags the gaseous disk to a greater extent than the stellar disk is likely robust given the supporting results from both the on-axis sample and full sample of RGB stars. Although the MDF of the mixed population is similar to the disk (Section~\ref{sec:mdfs}),\edit1{\footnote{RGB stars with uncertain disk/halo classficiations are more likely to be located away from the major axis (Section~\ref{sec:mdfs}). For on-axis stars, the disk and mixed MDFs are consistent for \fehphot\ $\gtrsim$ $-1.5$, where the mixed MDF contains fewer metal-poor stars.}} it is not immediately clear whether its AD is disk-like or halo-like. We favor the latter interpretation based on the more precise AD distribution from on-axis RGB stars, which may suggest that the mixed population is dominated by disk stars on kinematically hotter orbits (Section~\ref{sec:thick_disk}).

The metal-rich group of halo stars (\fehphot\ $\sim$ $-0.10$) identified in Section~\ref{sec:mdfs} has a median AD of \admedhalomr\ \kms\ based on RGB tracers on the major axis and a median AD of \admedhalomrall\ \kms\ based on all RGB tracers. In Figure~\ref{fig:ad_vs_vhelio}, this metal-rich halo group  corresponds to stars clustered near \vhelio\ $\sim$ $-280$ \kms\ with \pdisk\ $<$ 0.25. \edit1{The rest of the halo ($p_{\rm mr} < 0.75$) has ADs of \admedhalomp\ and \admedhalompall\ based on on-axis and all RGB tracers respectively.}
For both RGB samples, the AD of the metal-rich halo group 
is consistent with the rest of the halo within \adhalomrdiffsigrest$\sigma$. For the full RGB sample, the metal-rich halo group AD is 
\edit1{marginally consistent with the disk at \adhalomrdiffsigdiskall$\sigma$, but is more distinct at \adhalomrdiffsigdiskonaxis$\sigma$ when comparing median ADs computed from the on-axis RGB sample.}
However, we again interpret the on-axis ADs as being more accurate. 

Thus, regardless of its disk-like metallicity, the metal-rich halo group has kinematics inconsistent with a stellar disk. We also note that despite the minor yet meaningful differences in AD, the RGB ADs are remarkably similar as a whole. This indicates that the stellar disk is almost as removed from the gaseous disk as the stellar halo, providing evidence that the disk has experienced significant dynamical disturbance(s) (Section~\ref{sec:kicked_up}).
 

\subsection{Radial Metallicity Gradients}
\label{sec:grad}

\begin{figure}
    \centering
    \includegraphics[width=\columnwidth]{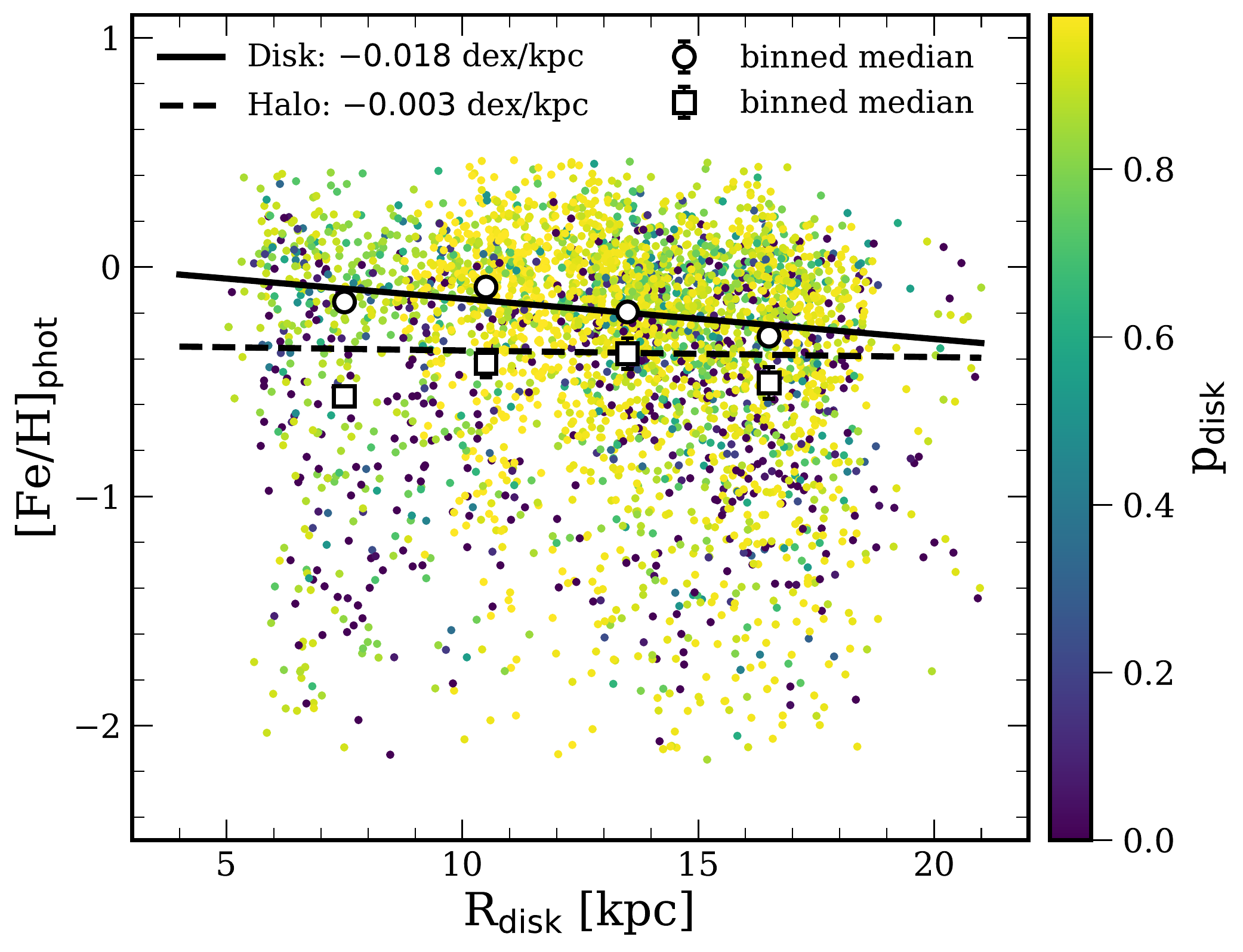}
    \caption{\fehphot\ as a function of de-projected radius in the disk plane (R$_{\rm disk}$) for SPLASH RGB stars (Section~\ref{sec:grad}). The points are color-coded by disk probability. 
    We determined metallicity gradients by fitting to all stars in the disk (\pdisk\ $>$ 0.75; solid line) and halo (\pdisk\ $<$ 0.25; dashed line) populations. We also show the median \fehphot\ in radial bins for the disk (outlined circles) and halo (outlined squares).
    We find a gradient in the disk ($-0.018$ dex kpc$^{-1}$) and a weakly negative gradient in the halo.
    }
    \label{fig:fehphot_vs_rdisk}
\end{figure}

We investigated whether radial metallicity gradients are present in M31's disk and halo populations across the SPLASH survey region ($R_{\rm disk} \sim 6-18$ de-projected kpc). Based on photometry alone, M31's disk was previously found to possess a negative metallicity gradient ($-0.020 \pm 0.004$ dex kpc$^{-1}$) between $R_{\rm disk} \sim 4-20$ de-projected kpc from a sample of 7 million RGB stars in the PHAT survey \citep{Gregersen2015}. The authors accounted for photometric effects due to crowding and dust extinction, but not the effects of contamination from stars in the MW foreground or M31's halo. 

We measured the gradients using all stars in the disk (\pdisk\ $>$ 0.75) and halo (\pdisk\ $<$ 0.25) populations. We parameterized the gradients in terms of angles and transverse distances \citep{Hogg2010}, which we converted to traditional slopes and intercepts after sampling from the posterior probability distribution of the linear model \citep{Foreman-Mackey2013}.\footnote{We used 10$^{2}$ walkers, 10$^{3}$ steps, for a total of 5 $\times$ 10$^{4}$ samples from the latter 50\% of each chain. The final parameters are computed from the 16$^{\rm th}$, 50$^{\rm th}$, and 84$^{\rm th}$ percentiles of the posterior probability distributions.} Figure~\ref{fig:fehphot_vs_rdisk} shows the relationship between $R_{\rm disk}$ and \fehphot\ for RGB stars in SPLASH along with the radial gradients measured for the disk and halo. We found a slope of \gradientdisk\ $\pm$ \gradientdiskerr\ dex kpc$^{-1}$ for the disk, in agreement with \citet{Gregersen2015}, and a weak slope of \gradienthalo\ $\pm$ \gradienthaloerr\ dex kpc$^{-1}$ for the halo. The associated intercept is \fehphot\ = \interceptdisk\ $\pm$ \interceptdiskerr\ (\intercepthalo\ $\pm$ \intercepthaloerr) for the disk (halo). 

The treatment of dust extinction (Section~\ref{sec:phot_params}) has a minor effect on the measured gradients. Using the low-extinction RGB sample defined by \fred\ $\times$ \avdal\ $<$ 0.25 instead yields a slope of \gradientdisklowext\ $\pm$ \gradientdisklowexterr\ (\gradienthalolowext\ $\pm$ \gradienthalolowexterr) dex kpc$^{-1}$ and an intercept of \interceptdisklowext\ $\pm$ \interceptdisklowexterr\ (\intercepthalolowext\ $\pm$ \intercepthalolowexterr) for the disk (halo). Disregarding dust in M31's disk entirely yields a slope of \gradientdiskdust\ $\pm$ \gradientdiskdusterr\ (\gradienthalodust\ $\pm$ \gradienthalodusterr) and an intercept of \interceptdiskdust\ $\pm$ \interceptdiskdusterr\ (\intercepthalodust\ $\pm$ \intercepthalodusterr) for the disk (halo). Our main findings of a gradient of approximately $-$0.02 dex kpc$^{-1}$ in the disk and a weakly negative gradient in the halo are therefore robust against dust effects. We adopted \gradientdiskfid\ \edit1{(\gradienthalofid)} as an encompassing range for the disk \edit1{(halo)} gradient slope. 

In contrast to the slopes, the gradient intercepts are sensitive to the treatment of dust within 0.1 dex. Moreover, changes in the assumed stellar age result in absolute metallicity differences up to 0.26 (Section~\ref{sec:phot_params}). However, the fiducial isochrone age should not affect the gradient slopes owing to the relatively constant shape of the RGB with stellar age. Instead, \citet{Gregersen2015} found that the presence of a negative age gradient with a magnitude larger than 0.1 Gyr kpc$^{-1}$ could flatten an apparent metallicity gradient in M31's disk. Nonetheless, resolved PHAT-based star formation histories for M31's disk do not show notable variations in the stellar age distribution with spatial position \citep{Williams2015,Williams2017}.\footnote{The PHAT-based star formation histories have only a few age bins for stellar populations older than 5 Gyr, such that they cannot resolve age gradients at the 0.1 Gyr kpc$^{-1}$ level.} Moreover, observations of massive external disk galaxies suggest that M31's age gradient should be $\lesssim \abs{0.1}$ Gyr kpc$^{-1}$ (e.g., \citealt{califa,Goddard2017}).

We refer the reader to \citet{Gregersen2015} for a discussion of the metallicity gradient in M31's disk in the context of the literature. Recent developments regarding the thickness of M31's disk (J.~Dalcanton et al., in preparation) imply that this shallow radial gradient may be the result of merger-driven mixing combined with projection effects. In addition, populations of high-extinction, kinematically colder and low-extinction, kinematically hotter intermediate-age PNe along the line-of-sight to M31's disk have been recently found to have radial argon gradients of $-0.02$ dex kpc$^{-1}$ and 
$-0.005$ dex kpc$^{-1}$
\citep{Bhattacharya2022}, respectively, similar to the disk and halo RGB populations in this work. We also note that, although the halo metallicity gradient is weak over the scale of the probed disk region, it is similar to the photometric metallicity gradient previously measured for the halo along the minor axis (1 dex over 100 kpc; \citealt{Gilbert2014}).

\section{Discussion}
\label{sec:discuss}

\begin{figure*}
    \centering
    \includegraphics[width=\textwidth]{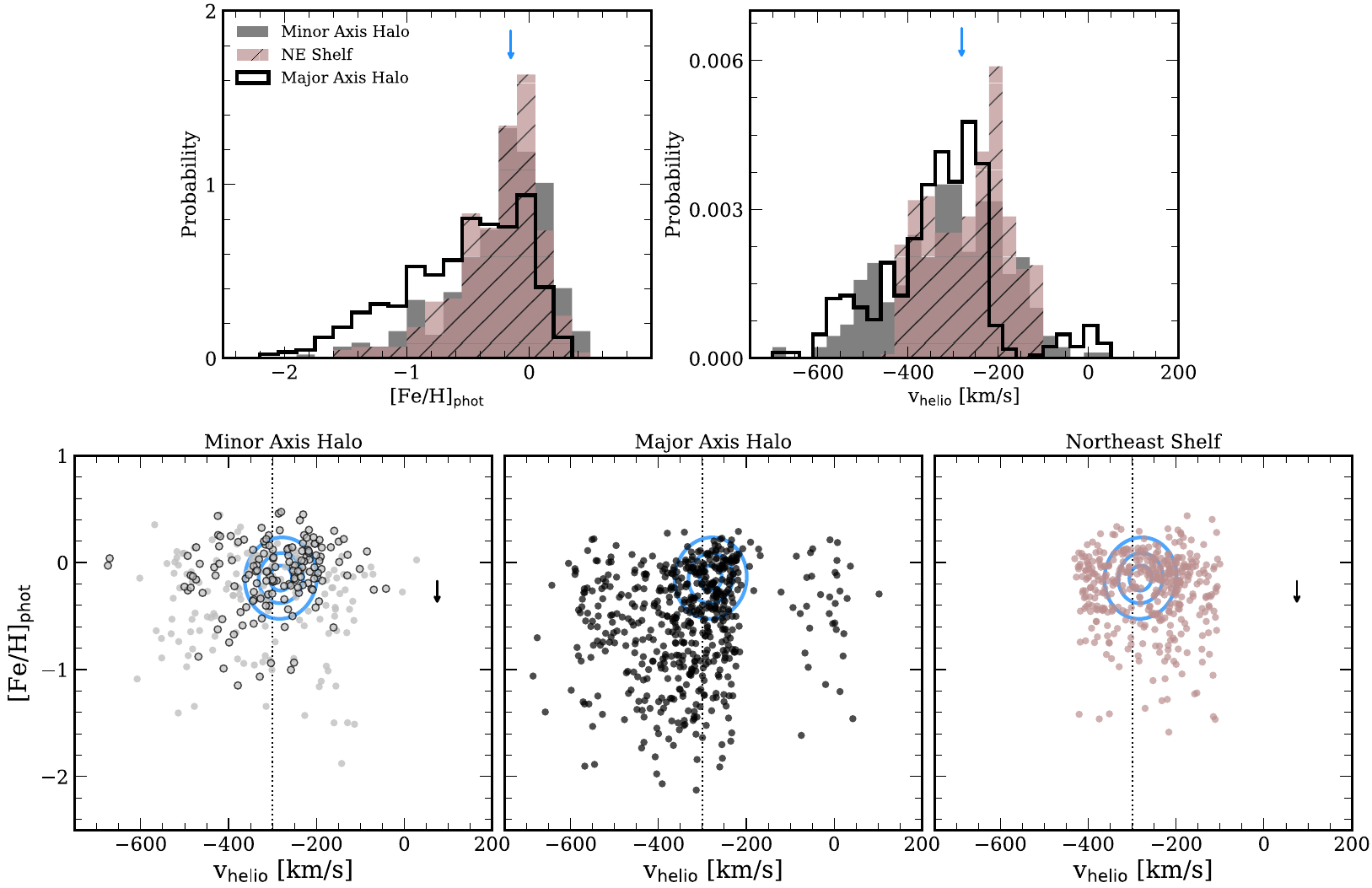}
    \caption{(Top) MDFs (left) and velocity distributions (right) for M31's stellar halo along the minor axis (gray histograms; \citealt{Gilbert2012,Gilbert2014}), the major axis (black outlined histograms; this work; \citealt{Dorman2012,Dorman2013,Dorman2015}), and for the NE shelf (brown hatched histograms; \citealt{Escala2022}).
    (Bottom) Photometric metallicity versus heliocentric velocity for the minor-axis halo (left), major-axis halo (middle), and NE shelf (right).
    See Section~\ref{sec:halo_comp} for the selection of the minor-axis halo and NE shelf samples and Section~\ref{sec:mdfs} for the major-axis halo. The outlined points (bottom left) are stars in the innermost minor axis field, {\tt f109}. The systemic velocity of M31 ($-300$ \kms) is shown as a dotted line. We assumed 4 Gyr isochrones to compare \fehphot\ between the three populations, though the mean stellar ages of the minor-axis halo and NE shelf are significantly older (Section~\ref{sec:halo_comp}). The black arrows in the bottom panels are the shift in \fehphot\ (\fehphotshiftfourtotwelve) when assuming 12 Gyr isochrones. Regardless of the assumed stellar age, the \fehphot\---\vhelio\ distribution of the major-axis halo is distinct from the minor-axis halo and NE shelf. In particular, the pronounced cluster of stars in the major-axis halo at \fehphot\ $\sim -0.10$ and \vhelio\ $\sim$ $-280$ \kms\ 
    is absent from the other populations  (top, blue arrows; bottom, blue contours; Section~\ref{sec:mdfs})
    }
    \label{fig:halo_comp}
\end{figure*}

In this section, we place our findings on the chemodynamical properties of the disk, halo, and mixed populations (Section~\ref{sec:results}) in a broader context. We discuss the nature of the mixed population and its implications for M31's thickened disk structure in Section~\ref{sec:thick_disk}. We show that the metal-rich group of halo stars (identified in Section~\ref{sec:mdfs}) is inconsistent with originating from the Giant Stellar Stream merger event in Section~\ref{sec:halo_comp}. We also demonstrate that M31's inner halo as probed along the major axis (i.e., near the disk) is distinct from the phase-mixed component previously studied along the minor axis (i.e., away from the disk). We put forward the hypothesis that this metal-rich halo group was kicked-up from the disk and that M31's inner halo possesses an in-situ population similar to the MW in Section~\ref{sec:kicked_up}.


\subsection{Disk Structure}
\label{sec:thick_disk}

The ``mixed'' population defined in Section~\ref{sec:disk_regions}, which constitutes \percentmixedsplash\% of the sample, consists of stars that cannot be securely associated with the stellar halo or dynamically colder disk. These stars have a disk-like MDF (Section~\ref{sec:mdfs}), but kinematical properties intermediate between the disk and halo (Section~\ref{sec:ad}). This raises the question of whether the mixed population is dominated by disk stars with uncertain kinematical classifications, potentially representing a ``thicker'' disk not captured by the adopted two-component velocity model (Section~\ref{sec:vmodel}). Alternatively, the mixed population could simply correspond to a transition region between a uniformly thick disk and the stellar halo, possibly including kicked-up disk stars (Section~\ref{sec:kicked_up}).

We found that modifying the line-of-sight velocity distribution models (Section~\ref{sec:disk_model}) to allow for a two-component disk does not result in better descriptions of the data as evaluated using Bayesian information criteria (Appendix~\ref{apdx:3comp}).
Currently, evidence in favor of a multiple-component disk structure in the PHAT region is lacking. Instead, the dominant RGB population in M31's disk region traces a thickened disk with a scale height of \diskscaleheight\ kpc as probed by the fraction of reddened stars (J.~Dalcanton et al., in preparation). This is substantially larger than the scale height of the integrated stellar disk of the MW (0.40 kpc; \citealt{BovyRix2013}) despite the relatively young age (4 Gyr; Section~\ref{sec:phot_params}) and high metallicity (\fehphotmeddiskrgb; Section~\ref{sec:mdfs}) of M31's disk. Moreover, the fundamentally thick nature of M31's disk is corroborated by its large line-of-sight velocity dispersion ($\sim$50-60 \kms) as traced by RGB stars (\citealt{Ibata2005,Dorman2012}; this work) and planetary nebulae (e.g., \citealt{Bhattacharya2019}). 

We note that the coarse relationship between velocity dispersion and metallicity found by \citet{Dorman2015} does not necessarily support a multiple-component disk structure, but rather likely reflects metallicity differences between spatial regions dominated by the stellar halo or disk. 
\edit1{This is because \citet{Dorman2015} did not separate disk and halo stars, but considered all RGB stars together, though their analysis is based on the same spectroscopic dataset as this work with similar membership criteria applied. For example, we found that separating RGB stars into metal-rich and metal-poor bins analogously to \citealt{Dorman2015} reproduces a negative correlation between metallicity and velocity dispersion, which is driven by the relative fractional contribution of halo stars to each metallicity bin.} Furthermore, the thick disk component argued for by \citet{Collins2011} in the outskirts ($\gtrsim$15 kpc) of M31's southern disk probably does not translate to the inner regions of M31's northern disk as probed in this work (see the discussion by J.~Dalcanton et al., in preparation).

We therefore favor the hypothesis that the mixed population mostly consists of stars from a thickened disk (with some halo contamination) that can be reasonably described by a single kinematical component. In general, an unambiguous kinematical detection of a multiple structural components in a projected disk would require comparisons to predicted line-of-sight velocity distributions obtained via forward modeling.

\subsection{The Minor-Axis Halo and Northeast Shelf}
\label{sec:halo_comp}

We compared stellar populations in M31's halo along its major axis (i.e., near the disk) to those previously studied along its minor axis (i.e., away from the disk). Our aim is to assess whether there is an evolutionary connection between the major-axis halo and disk or whether the major-axis and minor-axis halo share the same likely accretion-dominated origin (e.g., \citealt{Gilbert2014,McConnachie2018,Escala2020b}). We also compared the major-axis halo to the Northeast (NE) shelf, a tidal shell likely associated with the GSS (e.g., \citealt{Ferguson2002,Ferguson2005,Escala2022,Dey2022}), given predictions that it overlaps with the PHAT region and may therefore pollute the major-axis halo \citep{Fardal2007,Fardal2013}. The metal-rich nature of the GSS progenitor \citep{Gilbert2007,Gilbert2009,Ibata2007,Fardal2012,Gilbert2019,Escala2021,Escala2022} also raises the possibility that the metal-rich group of halo stars (Section~\ref{sec:mdfs}) could correspond to GSS-related tidal debris.

We used data for the minor-axis halo \citep{Gilbert2012,Gilbert2014} and NE shelf \citep{Escala2022} from SPLASH. We defined the minor-axis halo using spectroscopic fields spanning 9--18 projected kpc \citep{Gilbert2012},\footnote{This includes fields {\tt f109}, {\tt H11}, {\tt f116}, {\tt f115}, {\tt f207}, {\tt f135}, and {\tt f123}. We classified stars with likelihoods $\langle L_i \rangle > 0$ when including radial velocity as a diagnostic \citep{Gilbert2006} as M31 members.} which covers a radial range comparable to the disk region data (4--18 projected kpc). We also excluded known kinematically cold tidal debris from the Southeast shelf \citep{Gilbert2007} and GSS \citep{Kalirai2006,Gilbert2009} in the minor-axis halo by requiring that a star's probability of belonging to substructure is low ($p_{\rm sub} < 0.2$, where $p_{\rm sub}$ is defined analogously to \pdisk\ using the velocity models by \citealt{Gilbert2018}). For the NE shelf, we used the criterion $p_{\rm sub} > 0.75$, which excludes the majority of stars suspected to be disk contaminants \citep{Escala2022}.

To eliminate the \fehphot\ scale as a source of uncertainty in the comparison to the major-axis halo, we determined \fehphot\ homogeneously for the minor-axis halo and NE shelf using 4 Gyr isochrones (Section~\ref{sec:phot_params}). However, previously published \fehphot\ measurements from SPLASH for these stellar structures assume 12 Gyr ages, where the mean ages of the minor-axis halo and NE shelf are 10--11 Gyr \citep{Brown2007,Brown2008} and 8 Gyr \citep{Ferguson2005,Richardson2008}, respectively. Assuming 12 Gyr instead of 4 Gyr ages shifts the \fehphot\ distributions by \fehphotshiftfourtotwelve\ dex, where the median \fehphot\ = \medfehphotminhalo\ and \medfehphotneshelf\ for the minor-axis halo and NE shelf, respectively, in the 4 Gyr case. 

We note that whether the innermost minor-axis halo field ({\tt f109}) at \rproj\ = 9 kpc is included also impacts the associated \fehphot\ distribution, where the exclusion of this field changes the median \fehphot\ to \fehphotshiftnoinnerfield.  Although M31's extended disk reaches beyond $R_{\rm disk} \sim 40$ kpc \citep{Ibata2005} and is expected to have a line-of-sight velocity on the minor-axis equivalent to M31's systemic velocity ($-300$ \kms), the disk fraction is predicted to be $\lesssim$10\% at $R_{\rm disk} = 38$ kpc in {\tt f109} \citep{Guhathakurta2005}. Furthermore, no evidence of a kinematically cold disk feature has been detected in the velocity distribution of this field \citep{Gilbert2007,Escala2020a}, so we included {\tt f109} for a more accurate representation of the minor-axis halo \fehphot\ distribution over this radial range.\footnote{Field {\tt f109} may also contain Southeast shelf stars (i.e., GSS-related tidal material), but at a level not kinematically distinguishable from the phase-mixed halo \citep{Gilbert2007}.}

Figure~\ref{fig:halo_comp} shows \fehphot\ distributions, \vhelio\ distributions, and the relationship between \fehphot\ and \vhelio\ for the minor-axis halo, major-axis halo, and NE shelf. Regardless of the adopted age, the metallicity versus velocity distribution of the major-axis halo is distinct from the minor-axis halo and NE shelf. Performing 10$^{3}$ Anderson-Darling tests where the \fehphot\ measurements were perturbed by their (Gaussian) uncertainties yields that the major-axis halo is inconsistent with being drawn from the same distribution as the minor-axis halo or the NE shelf at the 0.1\% significance level with 95\% confidence.\footnote{When excluding field {\tt f109} from the minor-axis halo sample, the major-axis halo is distinct from the minor-axis halo at the 1.8\% significance level within this confidence interval.} Moreover, the pronounced cluster of major-axis halo stars at \fehphot\ $\sim -0.10$ and \vhelio\ $\sim$ $-280$ \kms\ is missing from the other stellar structures.
\edit1{Even for different locations in M31, the velocity of any metal-rich cluster with the same origin would remain close to M31's systemic velocity of $-300$ \kms, given that it corresponds to a dynamically hot population characterized by a lack of rotation relative to the disk (Section~\ref{sec:ad}).
}

Thus, M31's major-axis halo likely contains stellar populations absent from the phase-mixed component of the minor-axis halo and not dominated by GSS-related tidal debris. Even when restricting the major-axis halo population to stars with $p_{\rm mr} < 0.25$, it remains distinct from the minor-axis halo and NE shelf.
This is also the case when limiting the MDFs of each stellar structure to \fehphot\ $>$ $-1$, which corresponds to the RGB region in the CMD that should be entirely free of contamination by metal-rich helium burning stars (Section~\ref{sec:phot_params}).

\begin{figure}
    \centering
    \includegraphics[width=\columnwidth]{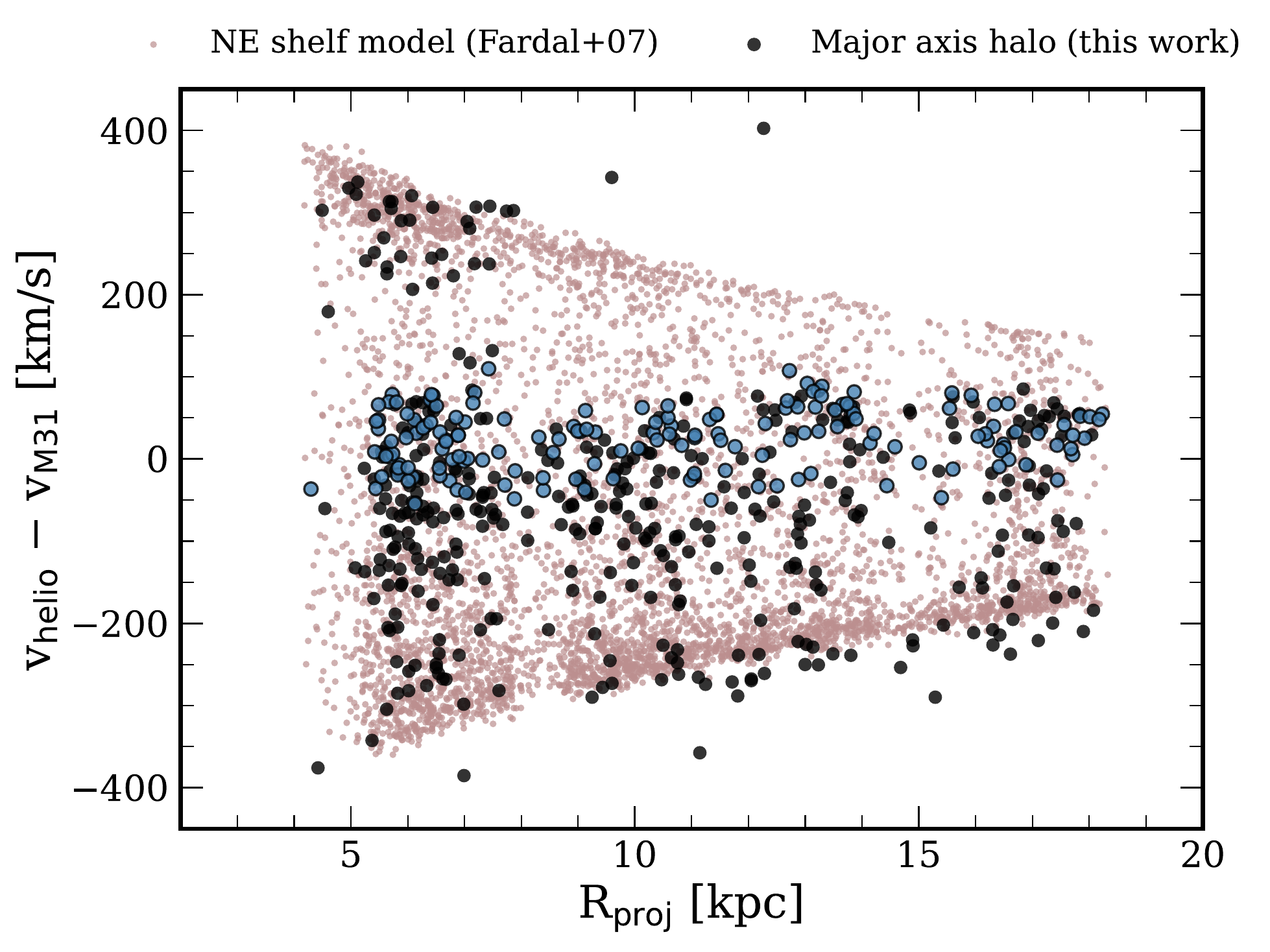}
    \caption{Heliocentric velocity shifted by M31's systemic velocity versus projected M31-centric radius for the major-axis halo (black circles; this work) compared against model predictions for NE shelf tidal debris (red points; \citealt{Fardal2007,Escala2022}) at the SPLASH data location (Figure~\ref{fig:fields}). Note that the model predicts M31 itself, not the NE shelf, to dominate the stars this region (Section~\ref{sec:halo_comp}). The metal-rich group of stars concentrated near \vhelio\ $-$ $v_{\rm M31}$ $\sim$ \edit1{20} \kms\ (blue circles; $p_{\rm mr} > 0.75$, Section~\ref{sec:ad}) is inconsistent with predictions for the NE shelf, which instead should be most visible at \vhelio\ $-$ $v_{\rm M31}$ $\lesssim$ $-$200 \kms\ in the SPLASH disk region \citep{Dorman2012}.
    }
    \label{fig:neshelf_model}
\end{figure}

We additionally evaluated whether such tidal material is likely to pollute the major-axis halo by comparing to predictions of N-body models for the formation of the GSS and NE shelf. We utilized a re-simulation of the \citet{Fardal2007} model for the complete disruption of a satellite progenitor 0.8 Gyr ago with stellar mass $M_{\rm sat} = 2.2 \times 10^9\ M_\odot$, which broadly provides a good match to observations of the NE shelf \citep{Escala2022}. Figure~\ref{fig:neshelf_model} shows the projected phase space distribution of the major-axis halo compared to the model predictions for NE shelf tidal debris at the location of the SPLASH disk fields (Figure~\ref{fig:fields}). The model predicts that M31 host particles constitute \frachostnbody\ of the stellar material in the SPLASH survey region, with the caveat that the specific fractional contribution depends on the assumed mass ($M_{\rm host} = 1.1 \times 10^{11} M_\odot$) and structural components of the host model. 
We therefore expect the NE shelf to be most detectable in the SPLASH region at velocities far removed from the disk (\vhelio\ $\sim$ $-$100 \kms) and along the high-density lower envelope of the tidal shell in projected phase space (\vhelio\ $\lesssim$ $-500$ \kms\ and \rproj\ $\sim 5-18$ kpc; Figure~\ref{fig:neshelf_model}; see also \citealt{Dorman2012}).\footnote{Interestingly, halo stars with \vhelio\ $<$ $-$500 \kms\ mostly have \fehphot\ $>$ $-1$ similar to the NE shelf (assuming 4 Gyr isochrones; Figure~\ref{fig:halo_comp}) and may have higher AD than the dominant halo population (Figure~\ref{fig:ad_vs_vhelio}). These properties could be consistent with a GSS-related origin.} Most stars in the major-axis halo, including the metal-rich group of stars concentrated near \vhelio\ $\sim -280$ \kms, are unlikely to originate from a disrupted satellite progenitor based on the expected debris pattern and the relative stellar density between M31's disk and the putative debris.

We have established that the major-axis halo over the SPLASH survey region probably does not predominately originate in the recent accretion of the GSS progenitor in a minor merger scenario. Detailed predictions for GSS-related tidal debris in M31's disk region in a major merger scenario \citep{Hammer2018,DSouzaBell2018} currently remain unexplored, although any such realistic simulation must be able to broadly reproduce the shell pattern of the NE shelf (similar to Figure~\ref{fig:neshelf_model}; \citealt{Escala2022,Dey2022}). The remaining halo formation channels involve ancient accretion and/or dissipative collapse or kinematical heating of the disk for an in-situ component. The dominant metal-poor halo population (\fehphot\ $\sim$ \fehphothalorgbmp; Section~\ref{sec:mdfs}) may have formed via accretion or in a ``classical'' in-situ scenario;  we discuss the possibility that the metal-rich group (\vhelio = \muvhalomr\ \kms, \fehphot\ = \muxhalomr) 
could have been kinematically heated from the disk by M31's last significant merger in Section~\ref{sec:kicked_up}.


\subsection{Kicked-Up Disk Stars}
\label{sec:kicked_up}

We have identified a metal-rich group of stars (\vhelio = \muvhalomr\ \kms, \fehphot\ = \muxhalomr, $f$ = \fmrhalo; Section~\ref{sec:mdfs}) in M31's major-axis halo that lags the rotation of the gaseous disk to a similar extent as the rest of the halo population (Section~\ref{sec:ad}). This group is inconsistent with observed and predicted properties for GSS-related tidal debris and appears to be absent from M31's minor-axis halo (Section~\ref{sec:halo_comp}). M31's bulge is too compact ($r_b = 0.78$ kpc; \citealt{Dorman2013}) to contribute to stellar populations in the surveyed disk region, such that M31's ``spheroid'' is indeed the stellar halo.

Based on a structural decomposition of M31's bulge, disk, and halo using I-band surface brightness profiles, PHAT lumnosity functions, and SPLASH kinematics, \citet{Dorman2013} found statistical evidence for an excess of stars (\fkickedupdiskdorman) that follow a disk-like luminosity function but have halo-like kinematics. We propose that the metal-rich group of halo stars identified in this work is the same ``kicked-up'' disk population. The primary difference between this work and \citet{Dorman2013} is that we performed a resolved \textit{chemodynamical} analysis that is minimally dependent on model assumptions. This has enabled us to characterize M31's putative kicked-up disk stars in detail \textit{as a distinct stellar population.} 
These stars contribute \fmrtot\ of the total RGB sample (see Appendix~\ref{apdx:contam} for the small effect of disk interlopers). 
Using the fractional contributions from Table~\ref{tab:vmodel}, the total statistical fraction of disk RGB stars in the sample is \fdisktotstat. The estimated kicked up disk fraction is therefore \fkickedupdiskexcess.

The chemical and kinematical properties of these metal-rich halo stars are broadly consistent with predictions of heated disk populations in similations of in-situ stellar halo formation. As noted by \citet{Dorman2013}, \citet{Purcell2010} found that $\sim$1\% of stars should be heated into the halo from the disk by a minor merger at a low impact angle, which is comparable to but systematically lower than estimates of the kicked-up disk fraction in M31. Moreover, stars heated from the disk can fractionally constitute $\sim$30\% of the inner stellar halo (e.g., \citealt{Tissera2013,Cooper2015,Khoperskov2022insitu}) as found in this work. Based on simulations including both major and minor mergers, \citealt{Jean-Baptiste2017} found that kicked-up disk populations can appear structured in kinematical phase space and should exhibit some degree of rotation inversely correlated with total accreted mass. Although other studies have similarly found that disk heated stars may show signs of rotational support, there can be significant halo-to-halo scatter (e.g., \citealt{McCarthy2012,Tissera2013}). M31's metal-rich clump of halo stars has a similar degree of rotational support as the rest of the dynamically hot halo population (Section~\ref{sec:ad}), where M31's inner halo slowly rotates \citep{Dorman2012}.

\subsubsection{Comparison to the MW}
\label{sec:mw}

In the MW, stars born in the proto-disk but kinematically heated onto high-eccentricity orbits by early merger(s) constitute the sole in-situ component of the stellar halo (e.g., \citealt{Bonaca2017,Haywood2018,DiMatteo2019,Belokurov2020}). In particular, the formation of in-situ halo, or ``Splash'', has been connected to the the Gaia-Enceladus-Sausage merger event (GES; e.g., \citealt{Helmi2018,Belokurov2018,Gallart2019,Bonaca2020,Grand2020}). This in-situ halo component has chemical abundances similar to the thick disk (e.g., \citealt{DiMatteo2019,Belokurov2020,Naidu2020}), old stellar ages comparable to the accreted component of the stellar halo (e.g., \citealt{Gallart2019,Bonaca2020}), and smoothly transitions in metallicity versus velocity space from the thin and thick disks (e.g., \citealt{Belokurov2020}). Furthermore, it fractionally constitutes $<$15\% of the MW's stellar halo for disk heights above 2 kpc \citep{Naidu2020}.

Although we cannot perform one-to-one comparisons with the MW's stellar halo, our results suggest that M31 may possess a fractionally larger (\fracpkickfid\ of halo RGB stars; Section~\ref{sec:mdfs}) in-situ halo component than the MW. In contrast to M31, the MDF of the MW's inner halo is strongly bimodal due to in-situ ([Fe/H] $\sim$ $-0.5$) and accreted ([Fe/H] $\sim$ $-1.2$) components dominated by the Splash and the GES merger remnant respectively (e.g., \citealt{Bonaca2017,DiMatteo2019}). However, given M31's more active merger history (e.g., \citealt{McConnachie2018}), there is no apriori reason to expect M31's inner halo to exhibit the same metallicity signatures as the MW. In simulations of stellar halo formation, the in-situ halo is generally more metal-rich than the accreted component \citep{Zolotov2009,Font2011,Tissera2012,Tissera2013,Tissera2014,Cooper2015,Pillepich2015,Khoperskov2022chem}, but otherwise halo MDFs can vary owing to scatter in formation histories.

\section{Summary}
\label{sec:summary}

We have combined optical HST photometry from PHAT with Keck/DEIMOS spectra from SPLASH to execute the first large-scale chemodynamical analysis of M31's inner disk region (4--19 kpc) based on metallicity and velocity measurements for \nrgbdustcorr\ RGB stars. We have performed a kinematical decomposition as a function of position across the disk region, where the line-of-sight velocity distributions are well-described by the combination of a thick stellar disk and stellar halo. As originally found by \citet{Dorman2012}, the disk-dominated (\percentdisksplash\% of RGB stars) region nevertheless has a substantial contribution from the inner stellar halo (\percenthalosplash\% of RGB stars) and an intermediate population with uncertain disk-halo classification (\percentmixedsplash\% of RGB stars). We have further found that:
\begin{enumerate}
    \item Assuming 4 Gyr stellar ages, the disk is characterized by a dominant metal-rich population (median \fehphot\ = \fehphotmeddiskrgb\ when corrected for dust effects; Section~\ref{sec:phot_params},~\ref{sec:mdfs}). The stellar halo is more metal-poor (\fehphot = \fehphotmedhalorgb), but contains a non-negligible fractional contribution ($f$ = \fmrhalo) from a stellar population with disk-like metallicity (\fehphot\ = \muxhalomr) that appears as a continuous extension of the disk in velocity space (\vhelio\ = \muvhalomr\ \kms).
    \item The AD, or rotational lag between the stellar and gaseous disks, is similar between the disk (\admeddisk\ \kms) and halo (\admedhalo\ \kms) populations (\citealt{Quirk2019}; Section~\ref{sec:ad}), suggesting that the disk has experienced significant dynamical heating. Despite this similarity, the halo AD is inconsistent with the disk, and the metal-rich halo stars have an AD (\admedhalomr\ \kms) consistent with the rest of the halo, suggesting that it does not correspond to a canonical thick disk.
    \item The disk metallicity gradient is \gradientdiskfid\ dex kpc$^{-1}$, in agreement with \citet{Gregersen2015}. This shallow gradient may originate from merger-driving mixing combined with projection effects. The halo metallicity gradient is 
    similar to that measured over 100 kpc scales along the minor axis \citep{Gilbert2014} at \gradienthalofid\ dex kpc$^{-1}$ 
    (Section~\ref{sec:grad}).
    \item RGB stars with uncertain disk-halo classifications have a disk-like MDF (Section~\ref{sec:mdfs}) and AD intermediate between the disk and halo (Section~\ref{sec:ad}). Rather than corresponding to a second thicker disk component, this mixed population is likely dominated by stars from a single thickened disk as it transitions into a stellar halo (Section~\ref{sec:thick_disk}).
    \item The MDF of M31's inner halo along the major axis (i.e., near the disk) is distinct from the halo MDF probed along the minor axis (i.e., away from the disk) over an equivalent radial range (Section~\ref{sec:halo_comp}). The metallicity and projected phase space properties of the metal-rich major-axis halo stars are also inconsistent with observations of the Northeast shelf and predictions for GSS-related tidal debris in a minor merger scenario (\citealt{Fardal2007,Escala2022}). This indicates that they were probably not accreted onto the halo.
    \item The chemical and kinematical properties of the metal-rich halo stars (Section~\ref{sec:mdfs},~\ref{sec:ad}) broadly agree with predictions of heated disk populations and some expectations for an in-situ halo based on current knowledge of the MW (Section~\ref{sec:kicked_up}). The estimated fraction of kicked-up disk stars is \fkickedupdiskexcess\ of RGB stars in the surveyed region, in agreement with the statistically inferred value of \fkickedupdiskdorman\ from \citet{Dorman2013}.
\end{enumerate}

These findings point to a scenario in which M31's inner stellar halo along the major-axis is distinct from the minor-axis halo, implying potentially disparate origins for each stellar structure. In particular, the minor-axis halo may be dominated by accretion from the GSS merger and more ancient events (e.g., \citealt{Brown2006,Gilbert2007,Gilbert2014,Ibata2014,McConnachie2018,Escala2020b,Dey2022}), whereas the metal-rich nature of the major-axis halo and the thickened nature of the disk suggests an entangled evolutionary history potentially driven by merger(s) and subsequent disk heating.

\acknowledgments

We thank the anonymous referee for a careful reading of this manuscript, which improved its clarity.
We also thank Mark Fardal for sharing the N-body model used in this work and Ekta Patel for helpful conversations regarding simulations.
IE acknowledges generous support from a Carnegie-Princeton Fellowship through the Carnegie Observatories and Princeton University. This material is based upon work supported by the NSF under Grants No.\ AST-1909759 (PG, ACNQ) and AST-1909066 (KMG). Some of the data presented in this article were obtained from the Mikulski Archive for Space Telescopes (MAST) at the Space Telescope Science Institute. The specific observations analyzed can be accessed via\dataset[10.17909/T91S30]{https://doi.org/10.17909/T91S30}.

We are grateful to the many people who have worked to make the Keck Telescope and its instruments a reality and to operate and maintain the Keck Observatory. The authors wish to recognize and acknowledge the very significant cultural role and reverence that the summit of Maunakea has always had within the indigenous Hawaiian community.  We are most fortunate to have the opportunity to conduct observations from this mountain.

\vspace{5mm}
\facilities{Keck (DEIMOS), HST (ACS)}

\software{astropy \citep{AstropyCollaboration2013,AstropyCollaboration2018}, emcee \citep{Foreman-Mackey2013}, matplotlib \citep{matplotlib}, numpy \citep{numpy}, scipy \citep{scipy}
}

\appendix

\section{Milky Way Foreground Contamination}
\label{apdx:mw}

\begin{figure}
    \centering
    \includegraphics[width=\textwidth]{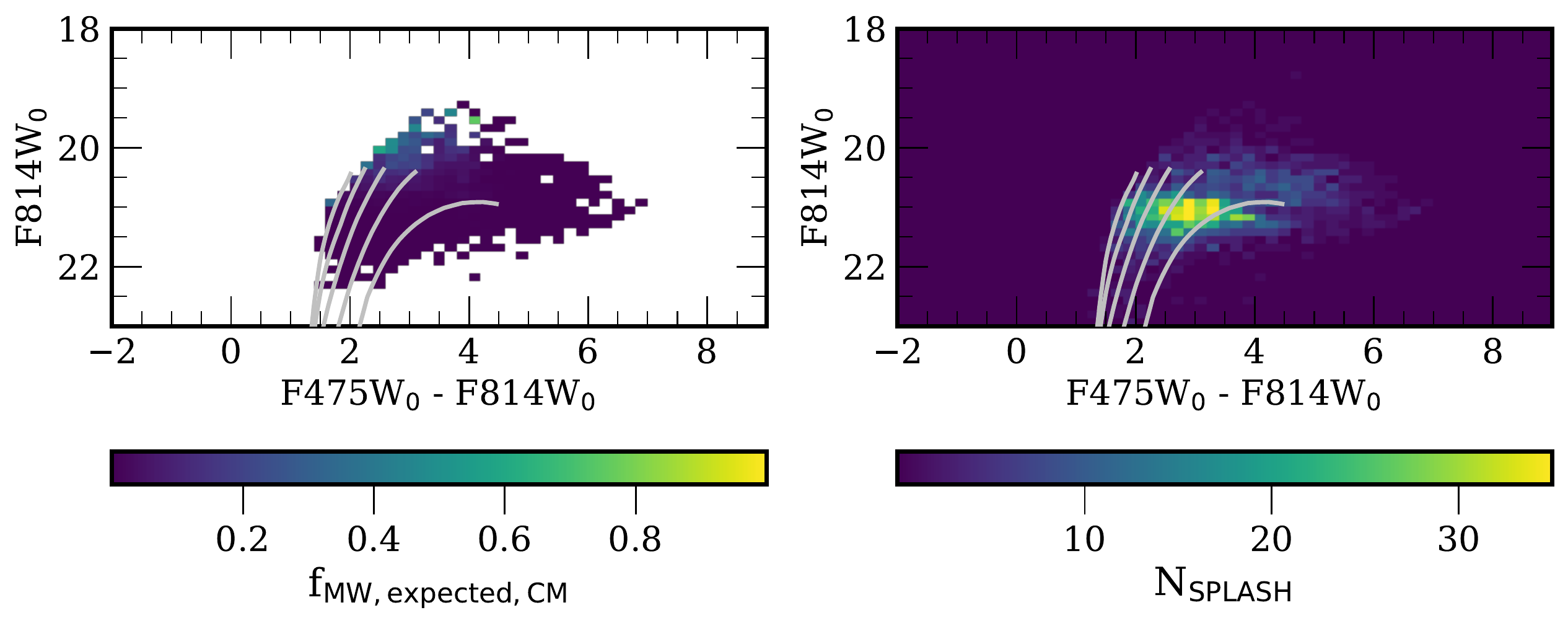}
    \caption{(Left) The fraction of expected MW contaminants (Eq.~\ref{eq:fcontam}) in the M31 giant star region of the CMD, defined using the CMD membership criterion from Section~\ref{sec:members}. We show 4 Gyr PARSEC RGB isochrones (gray lines) for reference as in Figure~\ref{fig:members}. The most contaminated portion of the CMD corresponds to relatively blue stars above the TRGB. (Right) The number density of M31 giant stars in SPLASH, classified using the CMD criterion (i.e., without \ewna\ measurements; Section~\ref{sec:members}). There are few SPLASH stars in the contaminated portion of the CMD (see also Figure~\ref{fig:members}). We expect $\sim$\fmwcontam\ of SPLASH stars to be MW contaminants based on CMD information alone.
    }
    \label{fig:contam}
\end{figure}

\edit1{As discussed in Section~\ref{sec:members}, we expect that contamination from the MW foreground along the line-of-sight to M31's disk is significantly less than in other regions of M31, such as its stellar halo along the minor axis, given that the stellar surface density of M31's disk is much higher than the MW. Here, we confirmed the negligible contribution from MW contaminants by using the Besan\c{c}on model\footnote{\url{https://model.obs-besancon.fr/modele_home.php}} \citep{Robin2003}. We simulated the MW foreground over the area and magnitude range spanned by the SPLASH survey, assuming all ages, spectral types, and distances out to 150 kpc for MW stars. We used photometric transformations from \citet{Sirianni2005} to convert the output BVRI photometry to the ACS system. We also refined the sample of MW foreground stars by limiting their location inside SPLASH fields on the sky (Figure~\ref{fig:fields}). We calculated the expected fraction of MW contaminants in a given color-magnitude (CM) bin,
\begin{equation}
  f_{\rm MW, expected, CM} = \frac{N_{\rm Besancon, CM}}{N_{\rm PHAT, CM}},
  \label{eq:fcontam}
\end{equation}
where $N_{\rm PHAT,CM}$ is the number of stars observed by the PHAT survey in each CMD bin. We applied the CMD criterion for M31 membership (Section~\ref{sec:members}) to restrict the MW foreground population to the CMD region corresponding to M31 giant stars. Figure~\ref{fig:contam} shows the expected fraction of MW contaminants ($f_{\rm MW, expected, CM}$; left panel) compared to the number density of SPLASH M31 giants (according to the CMD criterion) without \ewna\ measurements (right panel). The most contaminated portion of the CMD is F814W$_{\rm 0}$ $\lesssim$ 20.5 and (F475W $-$ F814W)$_{\rm 0}$ $\lesssim$ 4, where there are few SPLASH stars in this region (corresponding to relatively blue stars above the TRGB; see also Figure~\ref{fig:members}). We therefore expect that $\sim$\fmwcontam\ of SPLASH stars in the M31 giant portion of the CMD are MW contaminants based on CMD information alone (i.e., without taking into account spectroscopic diagnostics such as \ewna).}

\section{Velocity Models for Additional Regions}
\label{apdx:vmodel}

\begin{figure*}
    \centering
    \includegraphics[width=\textwidth]{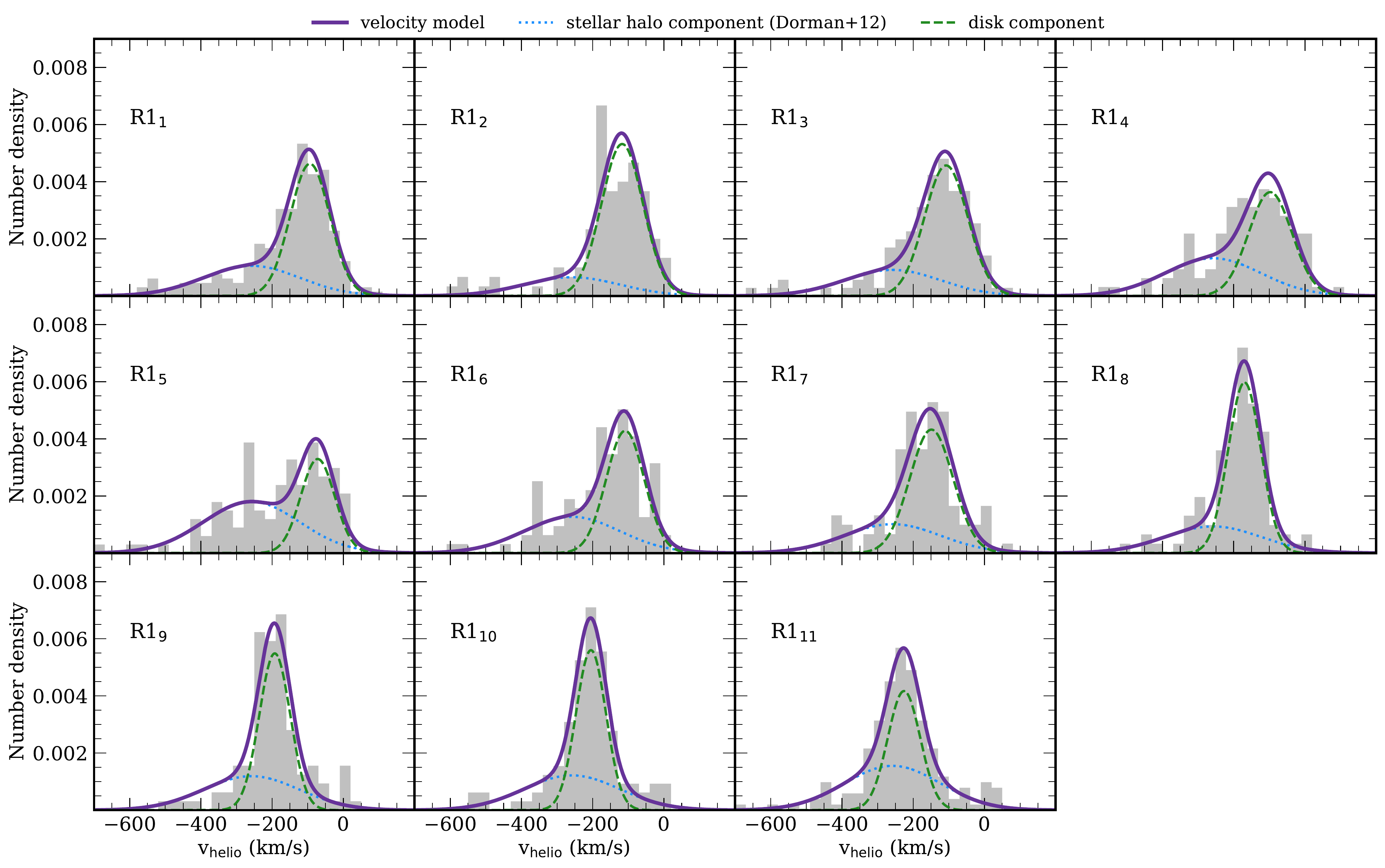}
    \caption{Same as Figure~\ref{fig:ne3}, except for region R1.}
    \label{fig:ne1}
\end{figure*}

\begin{figure*}
    \centering
    \includegraphics[width=\textwidth]{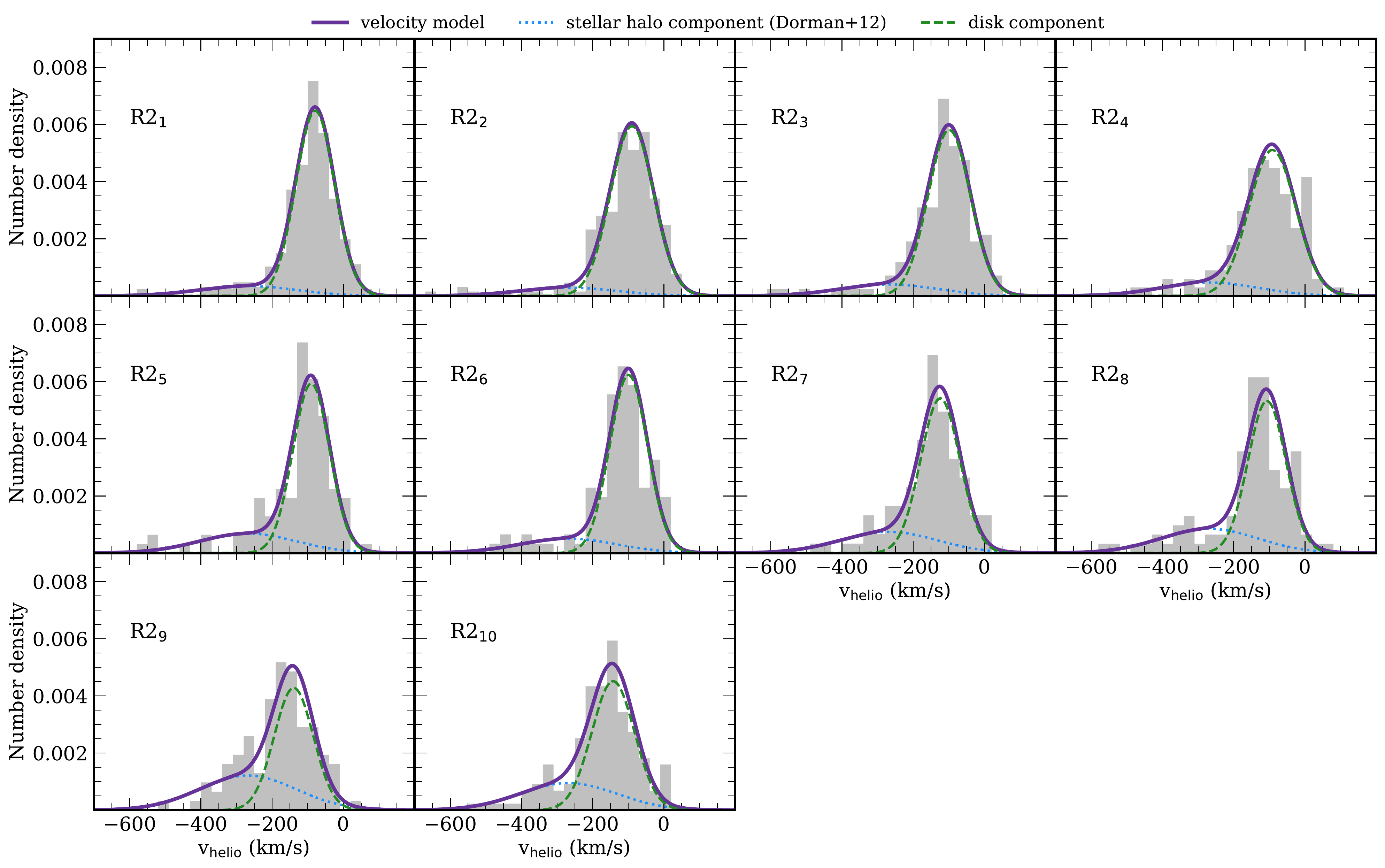}
    \caption{Same as Figure~\ref{fig:ne3}, except for region R2.}
    \label{fig:ne2}
\end{figure*}

Here, we show the velocity distributions and model fits (Section~\ref{sec:vmodel}) for regions R1 (Figure~\ref{fig:ne1}) and R2 (Figure~\ref{fig:ne2}) for completion.

\section{Infrared Color-Magnitude Diagram}
\label{apdx:dust}

\begin{figure}
\begin{minipage}[c]{0.48\textwidth}
    \centering
    \includegraphics[width=\textwidth]{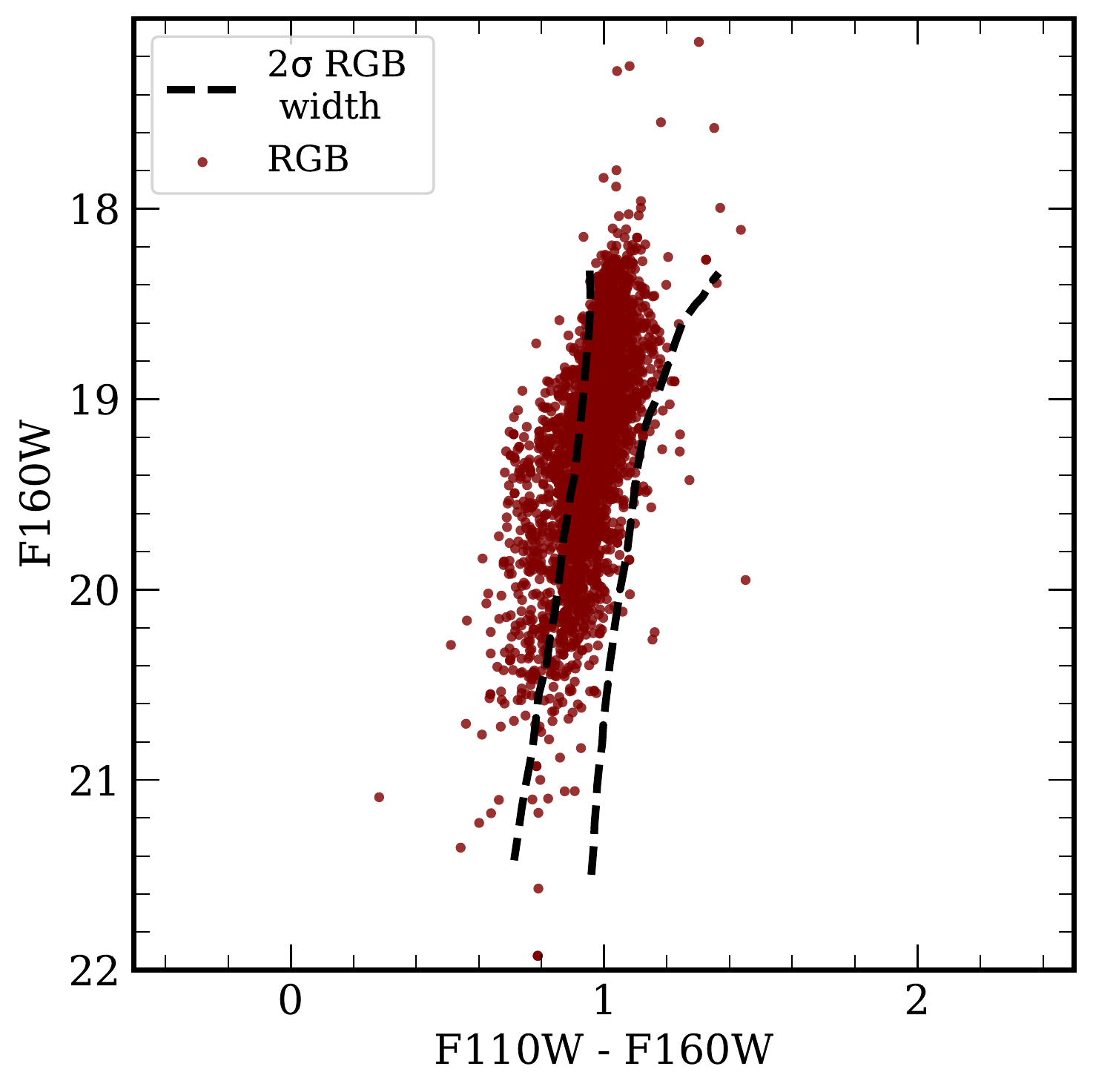}
    \caption{IR (F110W, F160W) PHAT v2 CMD \citep{Dalcanton2012,Williams2014} for RGB  
    stars in SPLASH (Appendix~\ref{apdx:dust}). The RGB classification is based on the foreground-reddening corrected optical CMD (Section~\ref{sec:zinit}). 
    The RGB stars follow a tight sequence insensitive to metallicity and age. Note that the axes are not shown to equal scale for clarity.
    The black dashed lines are the estimated 2$\sigma$ width expected for an unreddened RGB CMD, taking into account stellar surface density variations \citep{Dalcanton2015}.
    The lack of second broader sequence at redder colors suggests that the majority of stars are in front of the dust layer and not significantly reddened.
    }
        \label{fig:ir_cmd}
    \end{minipage}\hfill
    \begin{minipage}[c]{0.48\textwidth}
    \centering
    \includegraphics[width=\textwidth]{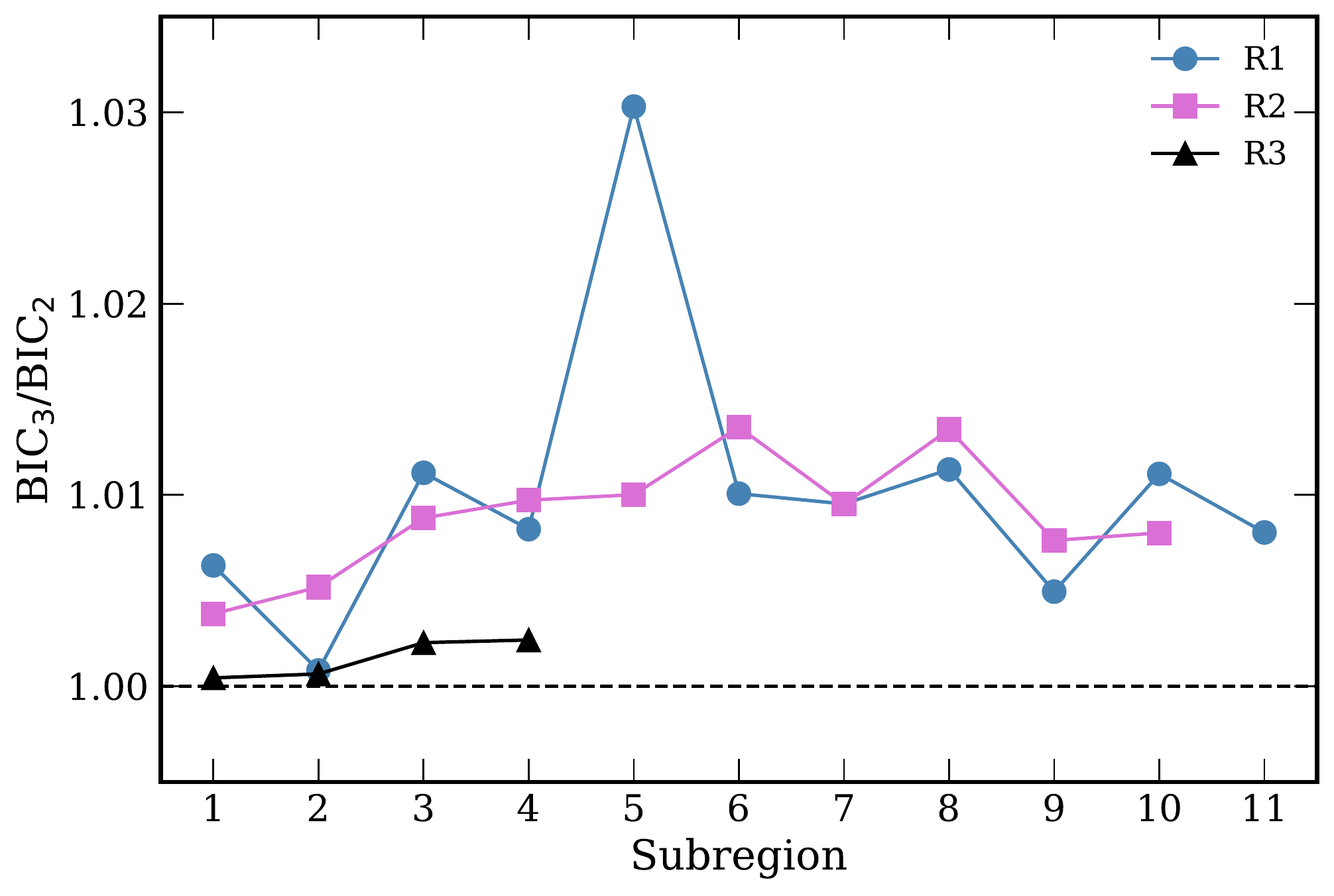}
    \caption{The ratio between the Bayesian information criterion (BIC) for the three-component and two-component velocity models for the line-of-sight velocity distribution of each subregion (Appendix~\ref{apdx:3comp}). The blue, magenta, and black lines correspond to radial regions R1, R2, and R3 respectively (Section~\ref{sec:disk_regions}). The dashed horizontal line is unity.
    For \edit1{all} subregions, M31's disk is well-described by a single thickened component (BIC$_2$ $<$ BIC$_3$). 
    Regardless, the similarity between the BIC values indicates that a simpler two-component model (halo, thick disk) is preferred. 
    }
    \label{fig:bic}
    \end{minipage}
\end{figure}

Figure~\ref{fig:ir_cmd} shows the observed PHAT v2 (F110W, F160W) CMD of RGB stars in SPLASH classified using the foreground-reddening corrected optical CMD (Section~\ref{sec:zinit}). The IR CMD consists of a tight sequence of unreddened stars, where this sequence is largely insensitive to age and metallicity variations, although it is weakly dependent on stellar surface density due to the effect of crowding on the photometry \citep{Dalcanton2015}. We note that the apparent excess of blue stars in Figure~\ref{fig:ir_cmd} is partly a consequence of the unequal axes scales. This blue population corresponds to metal-poor RGB stars (Section~\ref{sec:mdfs}) and a small number of red helium burning stars (Section~\ref{sec:zinit}).
The lack of a second broader sequence at redder colors in Figure~\ref{fig:ir_cmd} implies that the majority of RGB stars in SPLASH are in front of the thin dust layer of M31's disk, in agreement with results from the low-extinction MDF (Figure~\ref{fig:mdf_lowext}).

\section{Structural Population Contamination}
\label{apdx:contam}

The presence of non-negligible groupings of metal-poor and metal-rich stars in the disk and halo (Section~\ref{sec:mdfs}), respectively, raises the question of how much contamination from halo star interlopers is expected in the disk population ($f_{\rm contam}^{\rm disk}$) and vice versa for the halo population ($f_{\rm contam}^{\rm halo}$). We defined the halo (disk) populations using the threshold \pdisk\ $<$ 0.25 
 (\pdisk\ $>$ 0.75) as in Section~\ref{sec:disk_model}. We computed $f_{\rm contam}^{\rm disk}$ and $f_{\rm contam}^{\rm halo}$ using the halo and disk component velocity models for each subregion $s$ in radial region $r$ (Eq.~\ref{eq:vmodel}, Section~\ref{sec:vmodel}). 

The \pdisk\ $>$ 0.75 criterion for disk stars corresponds to a bounded velocity range within each subregion ([$v_{\rm lo, disk}$, $v_{\rm hi,disk}$]) and conversely the \pdisk\ $<$ 0.25 criterion for halo stars corresponds to a set of unbounded ranges at the extrema of a subregion's velocity distribution ([$-\infty$, $v_{\rm lo, halo}$] and [$v_{\rm hi,halo}$, $\infty$]). For each subregion, we calculated the expected ratio of halo stars to disk stars ($\mathcal{R}_{r,s}$) in the ``disk'' velocity range and ``halo'' velocity ranges by integrating the velocity models over each domain. That is, for the disk velocity range,
\begin{equation}
    \mathcal{R}_{r,s}^{\rm disk} = \frac{ 
    \int_{v_{\rm lo,disk}}^{v_{\rm hi,disk}}(1 - f_s) \mathcal{N}_r(v) dv 
    }
    {\int_{v_{\rm lo,disk}}^{v_{\rm hi, disk}} f_s \mathcal{N}_s(v) dv },
\end{equation}
where the velocity models are defined as in Eq.~\ref{eq:vmodel} given the parameters in Table~\ref{tab:vmodel}. Analagously for the halo velocity ranges,
\begin{equation}
     \mathcal{R}_{r,s}^{\rm halo} = \frac{ 
    \int_{-\infty}^{v_{\rm lo,halo}} (1 - f_s) \mathcal{N}_r(v) dv +
    \int_{v_{\rm hi,halo}}^{\infty} (1 - f_s) \mathcal{N}_r(v) dv
    }
    {\int_{-\infty}^{v_{\rm lo,halo}} f_s \mathcal{N}_s(v) dv +
    \int_{v_{\rm hi,halo}}^{\infty} f_s \mathcal{N}_s(v) dv
    }.
\end{equation}
Based on the number of stars in each subregion ($N_{r,s}$), we converted $\mathcal{R}_{r,s}^{\rm disk}$ to an expected number of halo stars, $N_{r,s,h}^{\rm disk}$, and analogously for $\mathcal{R}_{r,s}^{\rm halo}$. 
Thus, $f_{\rm contam}^{\rm disk}$ is given by the expected number of interlopers in each subregion summed over the survey footprint,
\begin{equation}
\sum_r \sum_s N_{r,s,h}^{\rm disk} / (N_{r,s,h}^{\rm disk} + (1 - \mathcal{R}_{r,s}^{\rm disk}) \times N_{r,s})
\end{equation}
where $f_{\rm contam}^{\rm halo}$ is the complementary formulation. This yields $f_{\rm contam}^{\rm disk}$ = \fcontamdisk\ and $f_{\rm contam}^{\rm halo}$ = \fcontamhalo.

These contamination fractions indicate that the metal-poor disk and metal-rich halo populations, which have fractional contributions to the disk and halo MDFs of \fehphotdiskrgbmpfrac\ and \fehphothalorgbmrfrac\ respectively, cannot be solely explained by interlopers originating from a predominantly metal-poor halo or metal-rich disk. This is particularly unlikely for the metal-poor disk population owing to the high density of metal-poor stars with disk-like velocities (Figure~\ref{fig:vhelio_vs_fehphot}).
We estimated the maximal effect of metal-rich disk (metal-poor halo) interlopers on the halo (disk) MDF by subtracting the approximate number of stars corresponding to $f_{\rm contam}^{\rm halo}$ ($f_{\rm contam}^{\rm disk}$) from the halo (disk) sample. The stars are probabilistically removed according to their metallicity and the metal-rich (metal-poor) component model of the disk (halo) MDF, which we assumed to represent a ``true'' metal-rich disk (metal-poor halo) population. We performed 10$^3$ iterations to obtain a distribution on each statistical quantity, finding that the median and mean on \fehphot\ becomes \fehphotmedhalorgbcontamcorr\ (\fehphotmeddiskrgbcontamcorr) and \fehphotavghalorgbcontamcorr\ (\fehphotavgdiskrgbcontamcorr) for the halo (disk) population (cf.\@ Table~\ref{tab:fehphot}). This corresponds a difference of $-0.08$ ($-0.05$) in the median (mean) metallicity of the halo, whereas the values for the disk are unchanged.

Figure~\ref{fig:mdfs} shows the effect on the fiducial halo (disk) MDF in the extremal case of contamination by disk (halo) interlopers. The halo MDF shape is affected more than the disk, although the halo population retains a significant contribution from stars with disk-like \fehphot. The fraction of stars assigned to the metal-rich halo component based on a simple likelihood ratio definition decreases from \fracmrhalofid\ to \fracmrhalocontamcorr\ for the fiducial and contamination-corrected halo MDFs, respectively. We therefore do not expect more than \fracmrhalocontambydisk\ of metal-rich halo stars to be disk interlopers. Moreover, the fraction of stars assigned to the metal-rich halo group using the criterion $p_{\rm mr} > 0.75$ decreases from \fracpkickfid\ to \fracpkickcontamcorr\ in the case of this comparison.

\section{Three-Component Velocity Models}
\label{apdx:3comp}

To test whether the line-of-sight velocity distributions provide statistical evidence in favor of a two-component disk structure in M31 (Section~\ref{sec:thick_disk}), we fit the velocity distributions for each subregion with a three-component model nominally composed of a halo, thick disk, and thin disk. The likelihood function in this case is therefore,
\begin{equation}
    \ln \mathcal{L} = \sum_{i=1}^{N_{r,s}} \ln \left( f_{s_1} \mathcal{N} (v_i | \mu_{s_1}, \tau_{s_1}^{-1}) + f_{s_2} \mathcal{N} (v_i | \mu_{s_2}, \tau_{s_2}^{-1}) + (1 - f_{s_1} - f_{s_2}) \mathcal{N} (v_i | \mu_{r}, \tau_{r}^{-1} \right),
\end{equation}
in comparison to Eq.~\ref{eq:vmodel}. We required that $\mu_{s_1} > \mu_{s_2}$, $\sigma_{s_1} < \sigma_{s_2}$, $f_{s_1} > f_{s_2}$ for primary thin and secondary thick disk components, but otherwise retained the same assumptions and procedure as in Section~\ref{sec:disk_model}. We evaluated the goodness-of-fit of the three-component and two-component velocity models for each subregion using the Bayesian information criterion (BIC; Figure~\ref{fig:bic}). For \edit1{each} subregion, we found that a two-component model consisting of a halo and single thickened disk component provides a better description of the data. 
Regardless, the similarity between the BIC values indicates that a simpler two-component model provides an adequate statistical description of the data compared to a three-component model.

\end{document}